\documentclass[12pt]{article}

\usepackage{epsfig,amsmath,amssymb,verbatim,mathrsfs}

\numberwithin{equation}{section}

\def\beq{\begin{eqnarray}}
\def\eeq{\end{eqnarray}}
\def\bea{\begin{eqnarray}}
\def\eea{\end{eqnarray}}

\def\tev{\, {\rm TeV}}
\def\gev{\, {\rm GeV}}
\def\mev{\, {\rm MeV}}

\newcommand{\gsim}{\lower.7ex\hbox{$\;\stackrel{\textstyle>}{\sim}\;$}}
\newcommand{\lsim}{\lower.7ex\hbox{$\;\stackrel{\textstyle<}{\sim}\;$}}

\def\stilde{\widetilde}

\newcommand{\newc}{\newcommand}
\newc{\Nc}{N_{c}}
\newc{\CG}{C_G}
\newc{\gp}{g'}
\newc{\stopi}{\stilde t_i}
\newc{\sboti}{\stilde b_i}
\newc{\staui}{\stilde \tau_i}
\newc{\stopj}{\stilde t_j}
\newc{\sbotj}{\stilde b_j}
\newc{\stauj}{\stilde \tau_j}
\newc{\stopI}{\stilde t_1}
\newc{\stopII}{\stilde t_2}
\newc{\sbotI}{\stilde b_1}
\newc{\sbotII}{\stilde b_2}
\newc{\stauI}{\stilde \tau_1}
\newc{\stauII}{\stilde \tau_2}
\newc{\sstop}{s_{t}}
\newc{\cstop}{c_{t}}
\newc{\ssbot}{s_{b}}
\newc{\csbot}{c_{b}}
\newc{\sstau}{s_{\tau}}
\newc{\cstau}{c_{\tau}}
\newc{\Sstop}{s_{2t}}
\newc{\Cstop}{c_{2t}}
\newc{\Ssbot}{s_{2b}}
\newc{\Csbot}{c_{2b}}
\newc{\Sstau}{s_{2\tau}}
\newc{\Cstau}{c_{2\tau}}
\newc{\salpha}{s_\alpha}
\newc{\calpha}{c_\alpha}
\newc{\Calpha}{c_{2\alpha}}
\newc{\Salpha}{s_{2\alpha}}
\newc{\sbetapm}{s_{\beta_\pm}}
\newc{\cbetapm}{c_{\beta_\pm}}
\newc{\Sbetapm}{s_{2 \beta_\pm}}
\newc{\Cbetapm}{c_{2 \beta_\pm}}
\newc{\sbetaO}{s_{\beta_0}}
\newc{\cbetaO}{c_{\beta_0}}
\newc{\SbetaO}{s_{2 \beta_0}}
\newc{\CbetaO}{c_{2 \beta_0}}
\newc{\vu}{v_u}
\newc{\vd}{v_d}
\newc{\seL}{\stilde e_L}
\newc{\smuL}{\stilde \mu_L}
\newc{\seR}{\stilde e_R}
\newc{\smuR}{\stilde \mu_R}
\newc{\suL}{\stilde u_L}
\newc{\sdL}{\stilde d_L}
\newc{\suR}{\stilde u_R}
\newc{\sdR}{\stilde d_R}
\newc{\scL}{\stilde c_L}
\newc{\ssL}{\stilde s_L}
\newc{\scR}{\stilde c_R}
\newc{\ssR}{\stilde s_R}
\newc{\snue}{\stilde \nu_e}
\newc{\snumu}{\stilde \nu_\mu}
\newc{\snutau}{\stilde \nu_\tau}
\newc{\Gpm}{G^\pm}
\newc{\Hpm}{H^\pm}
\newc{\FFbS}{\overline{FF}S}
\newc{\FFbV}{\overline{FF}V}
\newc{\FSS}{F_{SS}}
\newc{\FSSS}{F_{SSS}}
\newc{\FFFS}{F_{FFS}}
\newc{\FFFbS}{F_{\overline{FF}S}}
\newc{\FSSV}{F_{SSV}}
\newc{\FVS}{F_{VS}}
\newc{\FVVS}{F_{VVS}}
\newc{\FFFV}{F_{FFV}}
\newc{\FFFbV}{F_{\overline{FF}V}}
\newc{\Fgauge}{F_{\rm gauge}}
\newc{\DRbarprime}{$\overline{\rm DR}'$ }
\newc{\DRbar}{$\overline{\rm DR}$ }
\newc{\MSbar}{$\overline{\rm MS}$ }
\newc{\Yu}{{\bf Y}_u}
\newc{\Yd}{{\bf Y}_d}
\newc{\Ye}{{\bf Y}_e}
\newc{\Au}{{\bf a}_u}
\newc{\Ad}{{\bf a}_d}
\newc{\Ae}{{\bf a}_e}
\newc{\bm}{{\bf m}}
\newc{\zhol}{Z^{\rm hol}}

\newc{\rwino}{r_{\tilde W}}
\newc{\rmu}{r_{\tilde H}}
\newc{\ra}{r_A}
\newc{\ccdot}{\!\cdot\!}

\newcommand{\Phib}{\Phi_{-b}}
\newcommand{\phib}{\varphi_{-b}}
\newcommand{\phia}{\varphi_{a}}
\newcommand{\Phia}{\Phi_{a}}

\newcommand{\phio}{\varphi_{1}}
\newcommand{\phimo}{\varphi_{-1}}

\newcommand{\nnmb}{\nonumber}
\newcommand{\del}{\partial}

\newcommand{\lrf}[2]{\left(\frac{#1}{#2}\right)}
\newcommand{\ta}{\tilde{a}}

\newcommand{\ttt}{\tilde{t}}
\newcommand{\delfp}{\delta f_+}

\begin{document}

\setlength{\baselineskip}{0.2in}



\begin{titlepage}
\noindent
\begin{flushright}
MCTP-07-28  \\
CERN-PH-TH/2007-151
\end{flushright}
\vspace{1cm}

\begin{center}
  \begin{Large}
    \begin{bf}
Cosmic Strings from Supersymmetric Flat Directions\\
     \end{bf}
  \end{Large}
\end{center}
\vspace{0.2cm}

\begin{center}

\begin{large}
Yanou Cui$^a$, Stephen P. Martin$^b$, David E. Morrissey$^a$,
James D. Wells$^{c,a}$ \\
\end{large}
\vspace{0.3cm}
  \begin{it}
$^a$~Michigan Center for Theoretical Physics (MCTP) \\
Physics Department, University of Michigan, Ann Arbor, MI 48109
  \vspace{0.3cm}
\vspace{0.1cm}
\end{it}\\
  \begin{it}
$^b$~Physics Department, Northern Illinois University, DeKalb, IL 60115\\
Fermilab National Accelerator Laboratory, PO Box 500, Batavia IL 60510
\vspace{0.3cm}
\end{it}\\
\begin{it}
$^c$~CERN, Theory Division, CH-1211 Geneva 23, Switzerland
\vspace{0.1cm}
\end{it}\\

\end{center}

\center{\today}

\begin{abstract}

  Flat directions are a generic feature of the scalar potential
in supersymmetric gauge field theories.  
They can arise, for example, from $D$-terms associated with an 
extra abelian gauge symmetry.  Even when supersymmetry is
broken softly, there often remain directions in the scalar field
space along which the potential is almost flat.
Upon breaking a gauge symmetry along one of these almost flat directions, 
cosmic strings may form.  Relative to the standard cosmic string 
picture based on the abelian Higgs model, these flat-direction cosmic 
strings have the extreme Type-I properties of a thin gauge core 
surrounded by a much wider scalar field profile. We perform a 
comprehensive study of the microscopic, macroscopic, 
and observational characteristics of this class of strings.  
We find many differences from the standard string scenario, 
including stable higher winding mode strings,
the dynamical formation of higher mode strings from lower ones,
and a resultant multi-tension scaling string network in the early 
universe.  These strings are only moderately constrained by 
current observations, and their gravitational wave signatures 
may be detectable at future gravity wave detectors.
Furthermore, there is the interesting but speculative prospect
that the decays of cosmic string loops in the early universe
could be a source of ultra-high energy cosmic rays or non-thermal
dark matter.  We also compare the observational signatures of flat-direction
cosmic strings with those of ordinary cosmic strings as well as $(p,q)$
cosmic strings motivated by superstring theory.

\end{abstract}

\vspace{1cm}

\end{titlepage}

\setcounter{page}{2}

\tableofcontents

\vfill\eject



\section{Introduction}

  Cosmic strings are one-dimensional topological defects
that can be formed in the early universe~\cite{Nielsen:1973cs,
Hindmarsh:1994re,sbook}.  They are created if there is a
phase transition in which a $U(1)$ subgroup of a continuous
symmetry is broken.  Cosmic strings are stable because they carry
a conserved topological charge.  This charge is
integer-valued, corresponding to $\Pi_1(U(1)) = \mathbb{Z}$,
and is related to the number of times the phase of the $U(1)$ breaking
field winds at spatial infinity~\cite{Kibble:1976sj}.

  Unlike other types of topological defects, such as monopoles and
domain walls, cosmic strings can be formed at a wide range of
energy scales \emph{after inflation} without severely contradicting
the observed cosmology.  The generic problem with topological defects
is that, on account of their stability, they can easily come to
dominate the energy density of the universe~\cite{sbook}.
For cosmic strings there is an important loophole.
Topological stability only applies to infinitely long strings.
Cosmic string loops do not carry a net topological charge,
and they can decay into particle or gravitational radiation.
Such loops are formed when string segments intersect and exchange ends,
or \emph{reconnect} (or sometimes called \emph{intercommute}).
This allows a network of long cosmic strings
to regulate its energy by chopping itself up into loops which
radiate away.  Indeed, for a wide range of initial string densities,
analytic and numerical studies find that the competing processes
of string stretching (from the cosmic expansion) and loop
formation come to balance each other out.  The network evolves
towards a universal \emph{scaling solution} whose properties
are almost fully characterized by the cosmic string
tension~\cite{Bennett:1987vf,Allen:1990tv,
Martins:1995tg,Vincent:1996rb,Vanchurin:2005pa},
independent of the initial conditions.

  The vast majority of work on cosmic strings has focused on
the abelian Higgs model, in which a $U(1)$ gauge symmetry is
spontaneously broken by the condensation of a charged scalar field.
In this model, the vacuum expectation value~(VEV) of the complex
scalar field determines the mass of the gauge field, $m_V$,
and the physical scalar Higgs field, $m_S$, through the relations
\beq
m_V \simeq g\,v,~~~~~~~m_S \simeq \sqrt{\lambda}\,v,
\eeq
where $g$ is the gauge coupling, $\lambda$ is the scalar quartic self-coupling,
and $v$ is the VEV of the scalar.  The relative size of $m_V$ 
and $m_S$ determines how the strings interact.  For $m_V < m_S$, parallel
strings tend to repel at large distances while anti-parallel
strings attract~\cite{sbook}.
These strings are said to be Type-II, in analogy with superconductors.
When $m_V > m_S$, the strings attract for any relative
orientation, and they are said to be Type-I.  The attractive force
between parallel Type-I strings allows them to
form stable higher-winding modes.

  In most field theories, including the abelian Higgs model,
the masses $m_V$ and $m_S$ are naturally of the same order.  Much of the
previous work on cosmic strings has therefore dealt with Type-II
or weakly Type-I strings.  In the present work,
we will instead investigate the behavior of very strongly Type-I cosmic
strings, corresponding to $m_V \gg m_S$.
Our motivation to consider the extreme Type-I limit comes from
supersymmetry~\cite{Martin:1997ns}.
As we will show below, there exist supersymmetric field theories
in which $m_V \gg m_S$ arises in a natural way when
a $U(1)$ gauge symmetry is broken along a flat-direction
of the scalar potential.  Supersymmetry is essential because it ensures
that quantum corrections do not lift the flat direction.

  The key ingredients in our construction, supersymmetry and a new
$U(1)$ gauge symmetry, are each well-motivated in their own right
independently of cosmic strings.
Low-energy supersymmetry is one of the most elegant ways to
explain the large hierarchy between the electroweak scale and
the Planck scale~\cite{Martin:1997ns}.  It can also provide a candidate
for the dark matter~(DM) in the lightest superpartner particle~(LSP),
and in its minimal form, leads to an excellent unification of
gauge couplings.  Supersymmetry also plays an important role in superstring
theories of gravity.  Additional local $U(1)$ symmetries arise in many
models of new physics such as grand unified models and $D$-brane
constructions~\cite{abelian factors}.
In supersymmetric models, such symmetries can also help to solve
the $\mu$ problem~\cite{muterm}.

  A common feature of supersymmetric theories is
the existence of directions in the scalar potential that are almost flat.
To be precise, an almost flat direction is one for which
the curvature of the potential near the minimum is much smaller than the scale
of the (symmetry-breaking) VEV at that minimum.  Typically, these
directions in field space are completely flat at tree-level,
when only renormalizable operators are included in the potential,
but they are lifted by higher-dimensional operators, quantum effects,
and supersymmetry breaking.  As long as the supersymmetry breaking
effects are both soft and small, the residual approximate supersymmetry
prevents quantum corrections from destroying the flatness of the potential.
When a $U(1)$ gauge symmetry is broken along an almost-flat direction,
the scalar excitation around the VEV along the flat direction is
much lighter than the corresponding massive gauge boson.
We will show that the cosmic strings
associated with this pattern of gauge symmetry breaking are of the strongly
Type-I sort~\cite{Hill:1987qx,Freese:1995vp,Barreiro:1996dx,Penin:1996si,
Perkins:1998re,Davis:2005jf,Donaire:2005qm}.\footnote{
Let us also emphasize that the cosmic strings arising in general 
(approximately) supersymmetric theories need not be associated with 
a flat direction, and can also be of the Type-II variety.  
For examples, see Refs.~\cite{Davis:1997bs,Davis:1997ny,
Endo:2003fr,Jeannerot:2003qv}.}

  The interactions and cosmological consequences of strongly Type-I
strings can be qualitatively different from those of Type-II and
weakly Type-I strings~\cite{Laguna:1989hn,Bettencourt:1994kc,
Bettencourt:1996qe}.  
When Type-I or Type-II cosmic strings intersect,
they can reconnect or pass through each other.  There is a third
possible outcome when a pair of strongly Type-I strings intersect.
Due to their mutual attraction,
two strong Type-I strings with topological charges $N_1$ and $N_2$
can combine to form a single stable string with topological charge
$N_{zip} = (N_1+ N_2)$ or $N_{zip} = |N_1-N_2|$.
%
%
At the point of intersection,
the incident strings can coalesce into a single higher-winding string,
which may then proceed to grow like a zipper~\cite{Bettencourt:1994kc}.
If this growth continues indefinitely, the outcome will be a single
higher-winding mode string of horizon length.  For Type-II
and weakly Type-I strings, previous calculations and simulations
predict that the outcome of a string
intersection is reconnection with a probability close to unity,
$P_r \simeq 1$~\cite{Copeland:1986ng,Shellard:1987bv,Matzner:1988,
Achucarro:2006es}.  Since reconnection is essential
to the formation of string loops, which in turn are essential for
the strings to be cosmologically viable, deviations away from $P_r\simeq 1$
can significantly alter the picture of cosmic strings in the early universe.
In particular, if string zippering is common,
there can exist a stable population of higher winding mode
strings as well~\cite{Tye:2005fn,Copeland:2005cy,Leblond:2007tf,
Avgoustidis:2007aa}.

  Many of the exotic properties exhibited by the strongly Type-I
cosmic strings arising from supersymmetric flat directions
are also found in the $(p,q)$ cosmic strings emerging
from superstring 
theory~\cite{Polchinski:2004ia,Jones:2002cv,Dvali:2003zj,
Copeland:2003bj,Polchinski:2004hb,
Jackson:2004zg,Copeland:2004iv,Hanany:2005bc},
consisting of $p$ fundamental $F$-strings and $q$ $D$-strings.
These cosmic superstrings can merge to form the equivalent
of higher winding modes.  In many cases they also have reconnection
probabilities much less than unity, $P_r \lesssim 1$.  However,
flat-direction strings differ greatly from these $(p,q)$ strings
in their microscopic properties.  This is borne out in the
the relationship between the (effective) topological charge
and the string tension, as well as in the selection rules for
string zippering.  It may therefore be possible to distinguish
$(p,q)$ strings from flat-direction strings with the observation
of several string lensing events, each with a different apparent
relative value for the string tension.

  In the present work we study the properties and implications
of cosmic strings derived from the breakdown of a $U(1)$ gauge symmetry
along a supersymmetric flat direction.  We begin in Section~\ref{internal}
by studying the internal structure of flat-direction strings.
Here, we present a simple toy model for the flat direction
breaking, and we investigate approximate solutions to the equations
of motion and  study the string tensions using variational methods.
In Section~\ref{interact} we discuss the interactions between
cosmic strings.  We apply these results in Section~\ref{ntwrk},
where we study the formation and evolution of flat-direction
string networks in the early universe.
The observational signatures produced by these networks will
be the subject of Section~\ref{stringcosmo}.
Finally, Section~\ref{disc} is reserved for our conclusions.

  Several earlier papers have investigated cosmic strings associated with
flat-directions~\cite{Hill:1987qx,Freese:1995vp,Barreiro:1996dx,Penin:1996si,
Perkins:1998re,Davis:2005jf,Donaire:2005qm}.
These studies have predominantly focused on the lowest ($N=1$) winding mode.
We expand on these studies by exhibiting an explicit and natural
field theory model for the strings, and by discussing the new features
that arise from the existence of stable higher ($N>1$) winding modes.
These modes significantly alter the cosmological picture of the strings.


\section{String Profiles and Tensions\label{internal}}

  To begin, we introduce a simple class of models for 
a supersymmetric flat direction that could arise if there exists a $U(1)$
gauge group in addition to those contained in the 
minimal supersymmetric standard model~(MSSM).
Within these models, we study the cosmic string solutions
they support.  In particular, we find approximate solutions
to the classical equations of motion subject to the 
boundary conditions appropriate to a cosmic string, 
and we use these solutions to motivate a variational estimate 
of the string tension.  Even though we focus on a particular
class of models in the present section, we expect that many
of the qualitative features that we find are also applicable
to other cosmic string solutions associated with flat directions.

\subsection{$(a,b)$ Flat Directions}

  As a prototypical model for $U(1)$ symmetry breaking along
a supersymmetric flat direction, we consider the $(a,b)$ model
discussed in Ref.~\cite{Morrissey:2006xn}.  The model consists
of a supersymmetric $U(1)$ gauge theory containing chiral
superfields $\Phia$ and $\Phib$ with integer charges
$a$ and $-b$ respectively.  Except for the special case
$a=b=1$~\cite{Martin:1999hc},
we will assume that $a$ and $b$ are relatively prime with $a+b>3$.
Aside from the $(1,1)$ model, other charged fields must be present
in the theory for anomaly cancellation.  However, these will decouple
from the present discussion as long as they do not develop VEVs.

  When the charges $a$ and $b$ are relatively prime,
the leading superpotential operator built from $\Phia$ and $\Phib$ is
\beq
W \supset \frac{\lambda}{M_*^{a+b-3}}\Phia^b\Phib^a,
\eeq
where $M_*$ is a large mass scale above which our effective theory
breaks down.  We also include the soft supersymmetry
breaking operators
\beq
V_{soft} \supset -m_a^2|\phia|^2 - m_b^2|\phib|^2 - 
\left( \frac{A}{M_*^{a+b-3}}\phia^b\phib^a+h.c.\right),
\eeq
where $\phia$ and $\phib$ are the scalar component fields of 
$\Phia$ and $\Phib$, and $A$ is a dimension-one coupling on the 
order of the soft supersymmetry breaking scale,
$A \sim \sqrt{|m_a^2|}\sim \sqrt{|m_b^2|}$.\footnote{
A simple spurion analysis indicates that other, non-holomorphic
supersymmetry breaking terms from insertions in the
K\"ahler potential are subleading~\cite{Morrissey:2006xn}.}
In writing this expression, we have implicitly redefined the scalar
components of $\Phia$ and $\Phib$ such that $A$ is real and positive.
We have also taken the soft masses for $\phia$ and $\phib$ to be tachyonic.

  The leading contributions to the scalar potential in the model are therefore
\bea
V_F &=& \frac{|\lambda|^2}{M_*^{2a+2b-6}}\left(|b\,\phia^{b-1}\,\phib^a|^2
+|a\,\phia^b\,\phib^{a-1}|^2\right),\\
V_D &=& \frac{g^2}{2}\left(a|\phia|^2-b|\phib|^2\right)^2,\\
V_{soft} &=& -m_a^2|\phia|^2 - m_b^2|\phib|^2 
- \left(\frac{A}{M_*^{a+b-3}}\phia^b\phib^a +h.c.\right).
\eea
With $A$ real and positive, there will be a global minimum of the
potential with both $\phia$ and $\phib$ real and positive.
This minimum is unique up to gauge rotations.

  If $a+b>3$ the potential will be almost flat along the
$D$-flat direction defined by
\beq
a|\phia|^2 = b|\phib|^2.
\eeq
Along this direction, the potential is destabilized at the origin,
and is only restabilized at large field values by the higher dimensional
$F$ term operators.  Near the minimum, the excitation along the flat
direction is much lighter than the excitations orthogonal to it
as well as the gauge bosons.  This allows us to integrate out the
heavy modes and obtain an effective potential for the light
excitation.

  Let us restrict ourselves to the flat direction by setting
\beq
\phia = {v}\,\cos\alpha,~~~~~
\phib = {v}\,\sin\alpha,
\eeq
where
\beq
\cos\alpha = \sqrt{\frac{b}{a+b}},~~~~~\sin\alpha = \sqrt{\frac{a}{a+b}}.
\label{angles}
\eeq
The scalar potential for $v$ becomes
\beq
V(v) = -{P}\,v^2 - \lrf{2Q}{a+b}\,(v^2)^{(a+b)/2}
+ \lrf{R}{a+b-1}\;(v^2)^{a+b-1},
\eeq
with
\bea
P &=& \frac{b\,m_a^2+a\,m_b^2}{a+b},\nnmb\\
Q &=& \frac{A}{M_*^{a+b-3}}\left[\frac{a^ab^b}{(a+b)^{a+b-2}}\right]^{1/2},
\label{pqr}\\
R &=& \frac{|\lambda|^2}{M_*^{2a+2b-6}}\,
\left[\frac{a^ab^b}{(a+b)^{a+b-2}}\right]\,(a+b-1).\nnmb
\eea
In terms of these variables, the minimum is given by
\beq
v = \left[\frac{1}{2\,R}\left(Q + \sqrt{Q^2+4\,P\,R}\right)\right]^{1/(a+b-2)}.
\label{vmin}
\eeq
Parametrically, this is on the order of
\beq
v \sim \left({m}M_*^{a+b-3}\right)^{1/(a+b-2)},
\label{vminapprox}
\eeq
where $m$ is the generic soft mass.
Thus, we expect ${m} \ll v \ll M_*$.
The true minimum of the potential does not lie precisely along
the flat direction if $m_a^2 \neq m_b^2$.  However, the deviation
is very small, and can be expanded in powers
of $(m_a^2-m_b^2)/g^2v^2 \ll 1$.  We will discuss this further below.

  For the special $(1,1)$ case with field charges $\pm 1$,
we disallow the bilinear term as in Ref.~\cite{Martin:1999hc} and
\emph{only} include the next-to-leading order term in
the superpotential,
\beq
W_{(1,1)} =
\frac{\lambda}{M_*}\Phi_1^2\Phi_{-1}^2.
\eeq
The various terms in the potential are therefore
\bea
V_F &=& \frac{4|\lambda|^2}{M_*^2}\left(|\phio\phimo^2|^2
+ |\phio^2\phimo|^2\right),\\
V_D &=& \frac{g^2}{2}\left(|\phio|^2-|\phimo|^2\right)^2,\\
V_{soft} &=& -m_1^2|\phio|^2 - m_{-1}^2|\phimo|^2
- \left(\frac{A}{M_*}\,\phio^2\phimo^2 + h.c.\right).
\eea
In the following sections we will analyze in detail the equations
of motion resulting from this scenario.

\subsection{Equations of Motion and Approximate Solutions}

  The equations of motion for the system are
\bea
0 &=& D^{\mu}D_{\mu}\,\varphi_i + \frac{\del V}{\del\varphi_i^{*}}\;,\\
0 &=& \del_{\nu}F^{\nu}_{\phantom{\mu}\mu} + i\,g\,\sum_i\,Q_i\,(\varphi_i^*
\stackrel{\leftrightarrow}{D}_{\mu}\varphi_i)\,,
\eea
where $D_{\mu} = \del_{\mu} + i\,g\,Q\,A_{\mu}$ is the 
gauge-covariant derivative.

  To obtain an approximate solution to these equations that describes
a cosmic string, it is convenient to introduce an Ansatz for the vector
and scalar fields.  Our Ansatz for a string with winding number $N$ is
\bea
\phia &=& {v\,(1+\epsilon)}\,\cos\alpha\;
e^{iNa\phi}\,f_a(r),\nnmb\\
\phib &=& {v\,(1-\epsilon)}{}\,\sin\alpha\;
e^{-iNb\phi}\,f_b(r),
\label{ansatz}\\
A_{\phi} &=& \frac{N}{gr}\,\tilde{a}(r).\nnmb
\eea
In these expressions, $r$ and $\phi$ are the radial and angular
cylindrical coordinates relative to the string axis, $v$ is the
vacuum expectation value, and $\cos\alpha$ and $\sin\alpha$ are
defined in Eq.~\eqref{angles}.
The dimensionless parameter $\epsilon$ characterizes the deviation
from $D$-flatness at the absolute minimum, and will be treated as
a small number.
The functions $f_a(r)$, $f_b(r)$, and $\tilde{a}(r)$ are undetermined
string profiles.
They are subject to the boundary conditions
\beq
f_a,\;f_b,\;\ta \to 0~~\mbox{as}~~r\to 0,~~~~~~~~~
f_a,\;f_b,\;\ta \to 1~~\mbox{as}~~r\to \infty.
\label{bc}
\eeq
The relative winding numbers of $\phia$ and $\phib$
allow for both $D_{\phi}\phia$ and $D_{\phi}\phib$ to fall off more
quickly than $1/r$ as $r\to \infty$.  This is a necessary condition
for the string tension to be finite.

  Inserting the profile functions into the equations of motion,
we obtain
\bea
0&=& f_a''+\frac{1}{r}f_a'-\frac{N^2a^2}{r^2}\,(1-\ta)\,f_a
-{a}{}\lrf{ab}{a+b}
\left[(1+\epsilon)^2f_a^2-(1-\epsilon)^2f_b^2\right]\,f_a\label{faeq}\\
&&~~~~~- \frac{1}{v(1+\epsilon)\,c_{\alpha}}\,\frac{1}{g^2v^2}\,
e^{-iNa\phi}\frac{\del\tilde{V}}{\del\phia^*},\nnmb\\
0&=& f_b''+\frac{1}{r}f_b'-\frac{N^2b^2}{r^2}\,(1-\ta)\,f_b
+{b}{}\lrf{ab}{a+b}\left[(1+\epsilon)^2f_a^2
-(1-\epsilon)^2f_b^2\right]\,f_b\label{fbeq}\\
&&~~~~~- \frac{1}{v(1-\epsilon)\,s_\alpha}\,\frac{1}{g^2v^2}\,e^{iNb\phi}
\frac{\del\tilde{V}}{\del\phib^*},\nnmb\\
0&=& \ta'' - \frac{1}{r}\ta' + \lrf{2ab}{a+b}\left[
a\,(1+\epsilon)^2\,f_a^2+b\,(1-\epsilon)^2\,f_b^2\right]\,(1-\ta).
\label{aeq}
\eea
In these expressions we have separated out the $D$-term part
of the potential by defining $\tilde{V} = (V - V_D)$.
We have also written the cylindrical radial coordinate $r$ in units
of $(gv)^{-1}$.  Thus, when we discuss the value of $r$ in absolute terms,
it will always be relative to the scale $(gv)^{-1}$.
The equations of motion are complicated and non-linear,
but we can obtain approximate solutions in the three regions $r\ll 1$,
$1\ll r \ll gv/m$, and $r\gg gv/m$.  We consider each of these
regions in turn.

{\bf Region I: $r\ll 1$}

  For $r\ll 1$, we expect $f_a$, $f_b$, and $\ta$ to all be small.  
Expanding the equations of motion to linear order in the profiles, we find
\bea
f_a &\sim& r^{|Na|},\\
f_b &\sim& r^{|Nb|},~~~~~~~~~~~~(r\ll 1)\\
\ta &\sim& r^2,
\eea
This behavior agrees with the expectation from Refs.~\cite{
Hindmarsh:1994re,sbook}

{\bf Region II: $1\ll r\ll gv/m$}

  In the intermediate region $1\ll r \ll gv/m$, we expect
$f_a$, $f_b$, and $\ta$ to all be on the order of unity.
As we will discuss below, in this region it is also self-consistent to neglect
the contribution of $\tilde{V} = (V-V_D)$ to the equation of motion
and to set $\epsilon = 0$.  The equations of motion for $f_a$
and $f_b$ simplify if we rewrite them in terms of
$f_+(r)$ and $f_-(r)$, defined by
\beq
\left\{
\begin{array}{ccc}
f_+&=&\frac{1}{2}(f_a+f_b)\\
f_-&=&(f_a-f_b)
\end{array}\right.
\Leftrightarrow
\left\{
\begin{array}{ccc}
f_a&=& f_++\frac{1}{2}f_-\\
f_b&=& f_+-\frac{1}{2}f_-
\end{array}\right..
\eeq
The equations of motion for $f_a$ and $f_b$ then imply
\bea
0&\simeq& f_-''+\frac{1}{r}f_-'
-\lrf{2ab}{a+b}\left[(a+b)f_++\frac{1}{2}(a-b)f_-\right]f_+f_-,
\label{fmmid}\\
0&\simeq& f_+''+\frac{1}{r}f_+'- \frac{1}{2}\lrf{2ab}{a+b}\left[(a-b)f_+
+\frac{1}{2}(a+b)f_-\right]f_+f_-,
\label{fpmid}
\eea
As $r$ grows larger than unity, the boundary conditions imply
$f_+ \to 1$ and $f_-\to 0$.  If $f_+$ is slowly varying
in this region, the approximate solution for $f_-$ is
\beq
f_- \sim K_0(\sqrt{2ab}f_+r) \sim \sqrt{\frac{1}{r}}\,e^{-\sqrt{2ab}\,f_+\,r}.
\eeq
Thus, $f_-$ falls off quickly, corresponding to the damping of
the scalar excitation orthogonal to the flat direction.
With $f_-$ very small, the equation for $f_+$
reduces to
\beq
0\simeq f_+'' + \frac{1}{r}f_+'.
\eeq
The corresponding solution is
\beq
f_+ =f_0 \ln\lrf{r}{r_0},
\label{fpmidlog}
\eeq
for some constants $f_0$ and $r_0$.
Our approximate result is self-consistent because
$f_+$ is indeed a slowly-varying function of $r$.

  We can also use this expression for $f_+$ to check
the range of $r$ over which we can safely neglect
the effects of the $\tilde{V}$ term in the equation of motion.
For $f_- \ll 0$, $f_+ \sim 1$, this term is on the order of
$({m}^2/g^2v^2)\;f_+$, where $m$ is the scale of
the soft supersymmetry breaking terms.  The necessary condition for
ignoring the $\tilde{V}$ contribution to the equation of motion
to the level of approximation we are working to is
\beq
f_+'',~\frac{1}{r}f_+' \gg \lrf{m^2}{g^2v^2}\,f_+~~~~~
\Rightarrow~~~~~r\ll \frac{g\,v}{m}.
\eeq

  To track the evolution of the gauge profile it helps
to define $\delta \ta = 1 - \ta$.  The corresponding equation
of motion is
\beq
0\simeq \delta \ta'' -\frac{1}{r}\delta \ta' -
ab\,f_+^2\delta \ta,
\eeq
where we have made use of the fact that $f_-$ is expected to damp
out quickly and that $\epsilon \ll 1$.
The solution is
\beq
\delta \ta \propto r\,K_1\left(\sqrt{2ab}f_+r\right) \sim
\sqrt{r}\,e^{-\sqrt{2ab}f_+r}.
\eeq
Therefore, $\delta \ta$ is damped out exponentially as well,
and $\ta$ quickly approaches unity.  Let us point out that the physical
gauge boson mass is $\sqrt{2ab}\,g\,v$.  Thus, this mass controls the
width of the gauge field profile (remembering that $r$ is expressed
in units of $1/gv$ here), as well as the width of the profile of $f_-(r)$.

{\bf Region III: $r\gg gv/m$}

  In the very large field region, $r \gg gv/m$,
the flat potential $\tilde{V}$ and the deviation of $\epsilon$
from zero become relevant to the evolution of $f_+$ and $f_-$.
For these large values of $r$, it is convenient to write
\beq
\delfp = 1 - f_+,
\eeq
since we expect $|\delfp| \ll 1$.  Consider first the effect of
$\tilde{V}$ and $\epsilon \neq 0$ on the evolution of $f_-$.
The equation of motion to linear order in $f_-$ and $\delta f_+$ becomes
\beq
0&=& f_-''+\frac{1}{r}f_-'
- [2ab\,+\mathcal{O}(\epsilon)]\,f_-
- 4\,ab\,\epsilon + \lrf{m_a^2-m_b^2}{g^2v^2}.
\eeq
To be able to impose $f_- \to 0$, we must choose
\beq
\epsilon = \frac{1}{4\,ab}\lrf{m_a^2-m_b^2}{g^2v^2}.
\eeq
This is consistent with our previous assumption that $\epsilon \ll 1$.

  Inserting this value of $\epsilon$ into the linearized equation of motion
for $\delta f_+$, we find
\beq
0 = \delfp'' + \frac{1}{r}\delfp' - m_S^2\delfp,
\label{dfpeq}
\eeq
where $m_S^2$ is a positive constant on the order of $m^2/g^2v^2$.
In the units we are using, this is of the same size as the mass
of the light excitation about the almost-flat direction.
A possible constant term in Eq.~\eqref{dfpeq} vanishes through
the minimization condition for $v$ given in Eq.~\eqref{vmin}.
The solution for $\delfp$ in the very large $r$ region is therefore
\beq
\delfp  \propto K_0(m_S\,r) \simeq \sqrt{\frac{\pi}{2\,m_S\,r}}\,e^{-m_S\,r}.
\eeq
Again, this is consistent with the results of Refs.~\cite{
Hindmarsh:1994re,sbook}.

\subsection{String Tensions}

  Having obtained approximate expressions for the string profiles,
we  estimate the tension of cosmic strings
in the $(a,b)$ model for various values of the winding number $N$.
Using the Ansatz of Eq.~\eqref{ansatz}, the contributions
to the tension of a string in the $(a,b)$ model are
\bea
\mu_{rad}/\pi\,v^2 &=&
2\,\int_0^{\infty}dr\,r\,\left[\lrf{b}{a+b}(f_a')^2
+\lrf{a}{a+b}(f_b')^2\right],\label{muparts}\\
\mu_{ang}/\pi\,v^2 &=&
2\,N^2\,ab\,\int_0^{\infty}dr\,\frac{1}{r}
\left[\lrf{a}{a+b}\,f_a^2+\lrf{b}{a+b}\,f_b^2\right](1-\ta)^2,\nnmb\\
\mu_{mag}/\pi\,v^2 &=&
N^2\int_0^{\infty}dr\,\frac{1}{r}\,(\ta')^2,\nnmb\\
\mu_{pot}/\pi\,v^2 &=& \int_0^{\infty}dr\,r\,\frac{1}{g^2v^4}\,V(f_a,f_b).\nnmb
\eea

  Except near the origin, and certainly whenever the potential is relevant,
it is a very good approximation to set $f_a = f_b = f_+$.
In this limit, the potential can be written in the form
\beq
\frac{1}{g^2v^2}\,V(f) \simeq
-\delta_1\,(f_+^2-1) -\lrf{2}{a+b}\,\delta_2\,(f_+^{a+b}-1)
+\lrf{\delta_2+\delta_1}{a+b-1}\,(f_+^{2a+2b-2}-1).
\label{Vf equation}
\eeq
Here, we have implicitly assumed that $a+b\geq 4$.  The dimensionless
constants $\delta_1$ and $\delta_2$ are given by
\bea
\delta_1 &=& \frac{1}{a+b}\lrf{b\,m_a^2+a\,m_b^2}{g^2v^2},\label{deltas}\\
\delta_2 &=& \frac{1}{g^2v^2}\frac{A}{M_*^{a+b-3}}
\left[\frac{a^ab^b}{(a+b)^{a+b-2}}\right]^{1/2}\,v^{a+b-2}.\nnmb
\eea
Using the parametric value of the VEV given in Eq.~\eqref{vminapprox},
these constants are of size
\beq
\delta_{1,2} \sim \lrf{m}{M_*}^{2(a+b-3)/(a+b-2)}.
\eeq
For $M_* \sim M_{\rm Pl}$ and $m \sim \tev$,
we find $10^{-30} \lesssim \delta_{1,2} \lesssim 10^{-15}$.
Although the expressions presented above were formulated for
strings in the $(a,b)$ theory, they can also be applied to $(1,1)$ theory
cosmic strings.  The correct formulae for the $(1,1)$ case
are obtained by setting $a=b=1$ in the radial and angular components
of the tension ($\mu_{rad}$ and $\mu_{ang}$ in Eq.\,\eqref{muparts}), 
but $a+b=4$ in the expression for the potential (Eq.\,\eqref{Vf equation}).
This adjustment accounts for our inclusion of terms beyond the
leading order for $(1,1)$ strings.

  To estimate the string tensions, we have used variational methods
as in Ref.~\cite{Hill:1987qx}.  Our trial profile functions are inspired
by the approximate solutions found above.
They are
 \bea
f_a(r) &=& \left\{
\begin{array}{lr}
p_1(r/r_1)^{|Na|}&~~~~~r\leq r_1\\
p_5 + p_3\,\ln\lrf{r}{r_1}&
~~~~~r_1<r<r_2\\
1-p_4\,e^{-(r-r_2)/r_3}&r\geq r_2
\end{array}\right.\nnmb\\
f_b(r) &=& \left\{
\begin{array}{lr}
p_2(r/r_1)^{|Nb|}&~~~~~r\leq r_1\\
p_5 + p_3\,\ln\lrf{r}{r_1}&
~~~~~r_1<r<r_2\\
1-p_4\,e^{-(r-r_2)/r_3}&r\geq r_2
\end{array}\right.\label{profc}\\
\ta(r) &=& \left\{
\begin{array}{ll}
a_0\left[3\lrf{r}{r_a}^2-2\lrf{r}{r_a}^3\right]
&~~~r\leq r_a\\
1&~~~r>r_a
\end{array}\right..\nnmb
\eea
The undetermined parameters are
$\{r_1,\,r_2,\,r_3,\,r_a,\,p_1,\,p_2,\,p_3,\,p_4,\,p_5\}$.
We fix four of them, $p_1$, $p_2$, $p_3$, and $p_4$, by requiring continuity
at $r=r_1$ and $r_2$, and  differentiability at $r_2$ 
(where the solution is expected to be slowly varying) but not at $r_1$.

  For a $(1,1)$ model string with winding number $N = 1$ and
$\delta_1 = \delta_2 = 1\times 10^{-20}$, our variational estimate
of the tension is
\bea
\mu_{rad}/\pi v^2 &=& 0.09093,\nnmb\\
\mu_{ang}/\pi v^2 &=& 0.00247,\nnmb\\
\mu_{mag}/\pi v^2 &=& 0.00228,\label{mus}\\
\mu_{pot}/\pi v^2 &=& 0.00228,\nnmb\\
\mu_{tot}/\pi v^2 &=& 0.09796.\nnmb
\eea
The corresponding values of the variational parameters are
\bea
r_1 &=& 14.01,\nnmb\\
r_2 &\simeq& r_3 = 3.112\times 10^{9},\label{rs}\\
r_a &=& 36.26,\nnmb\\
p_5 &=& 0.04713.\nnmb
\eea
Recall that we express all dimensionful quantities in units of $1/gv$.

  As expected, the gauge profile is much narrower than the scalar profiles
({\it i.e.}, $r_a\ll r_2$), which have substantial support out to 
$r \sim 1/\sqrt{\delta_{1,2}}$.
The small $r$ power-law form of the scalar profiles extends out about as
far as the gauge profile ({\it i.e.}, $r_1\simeq r_a$), after which 
it continues to grow logarithmically
slowly until the profile reaches unity.  We also find that the total string
tension is dominated by the radial contribution.

  To a very good approximation, the shape of the profiles
and the value of the string tension do not depend on $\delta_1$ and
$\delta_2$ independently, but rather on the combination
\beq
\Delta = \delta_1+\delta_2/2.
\eeq
This can be seen explicitly by evaluating $\mu_{pot}$ using the
Ansatz profiles of Eq.~\eqref{profc} and keeping only the leading
terms in the expansion in $1/\ln(\delta_{1,2}) \ll 1$.

  We have investigated a number of other sets of profile functions
as well.  As long as the trial scalar profile increases sufficiently
(logarithmically) slowly in the region $1\ll r \ll 1/\sqrt{\delta}$
and drops rapidly for larger $r$, we find that the resulting
estimates for the string tension are very similar.
This gives us confidence that our estimates are 
close to the exact values.

\begin{figure}[ttt]
\vspace{1cm}
\begin{center}
        \includegraphics[width = 0.7\textwidth]{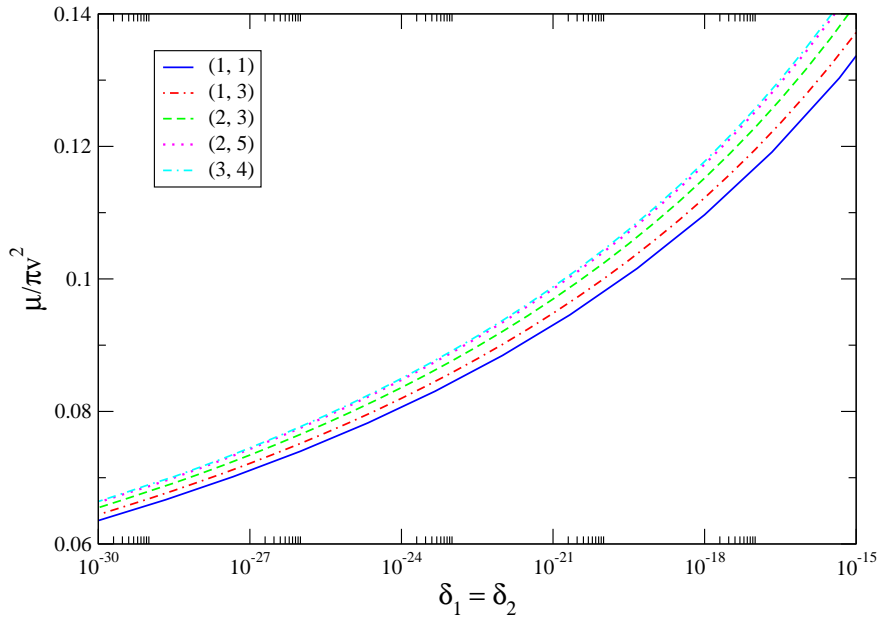}
\end{center}
\caption{Tensions of $N=1$ strings
as a function of the potential parameters $\delta_1=\delta_2$ 
for various $(a,b)$ theories. 
}
\label{mud2}
\end{figure}

 Fig.~\ref{mud2} shows the dependence of the string tension
in the $(1,1)$, $(1,3)$, $(2,3)$, $(2,5)$, and $(3,4)$ models
on the value of $\delta_1=\delta_2 = 2\Delta/3$ for a winding number $N=1$.
Even for very small values of $\delta_{1,2}$, corresponding to
extremely flat potentials, the string tension is within about
an order of magnitude of $v^2$.  Thus, while the string is
very wide in units of $1/g\,v$, the VEV still sets the size of the tension.
The tensions are also very similar for different values of $(a,b)$.
This is not very surprising given that the radial portion of the string
tension appears to be the dominant one.  In the $r\gg 1$ region,
we expect $f_a\simeq f_b$ so that the expression for the
radial contribution to the tension in Eq.~\eqref{muparts}
does not depend explicitly on $(a,b)$.  The dependence on $(a,b)$ only
then comes about through the size of the terms in the potential.
In generating Fig.~\ref{mud2}, we neglected this dependence by specifying
the value of $\delta_1=\delta_2$ explicitly.  Our results also suggest
that the detailed form of the (non-$D$) potential does not play a significant
role in determining the string tension or the string profiles other
than to set the scale at which the scalar profiles are cut off.

\begin{figure}[ttt]
\vspace{1cm}
\begin{center}
        \includegraphics[width = 0.7\textwidth]{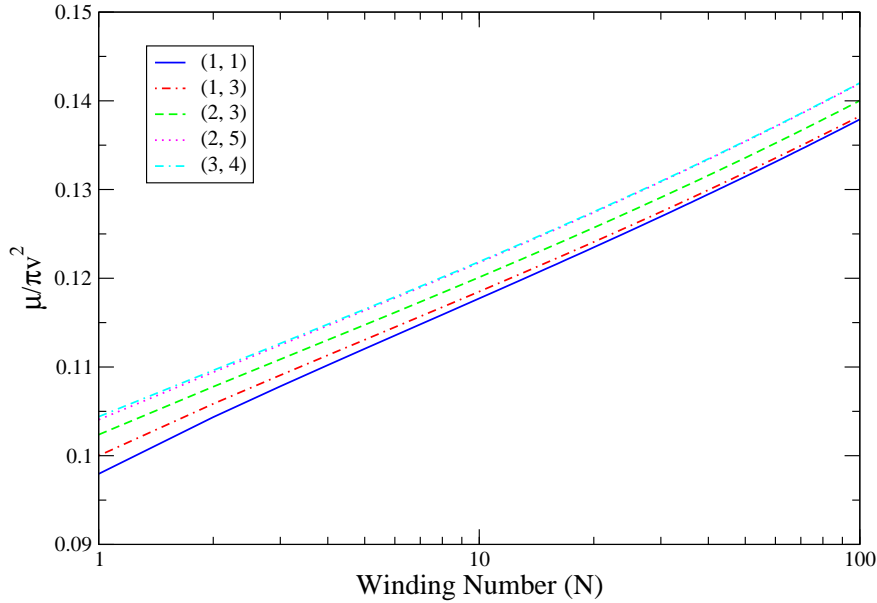}
\end{center}
\caption{String tensions 
as a function of the winding number $N$ for the
potential parameters $\delta_1=\delta_2 = 1\times 10^{-20}$ 
in various $(a,b)$ theories. 
Note that the tension of the $N=2$
string is much smaller than twice the tension of 
the $N=1$ string, thereby allowing stable $N=2$ strings.}
\label{munn}
\end{figure}

\begin{figure}[ttt]
\vspace{1cm}
\begin{center}
        \includegraphics[width = 0.7\textwidth]{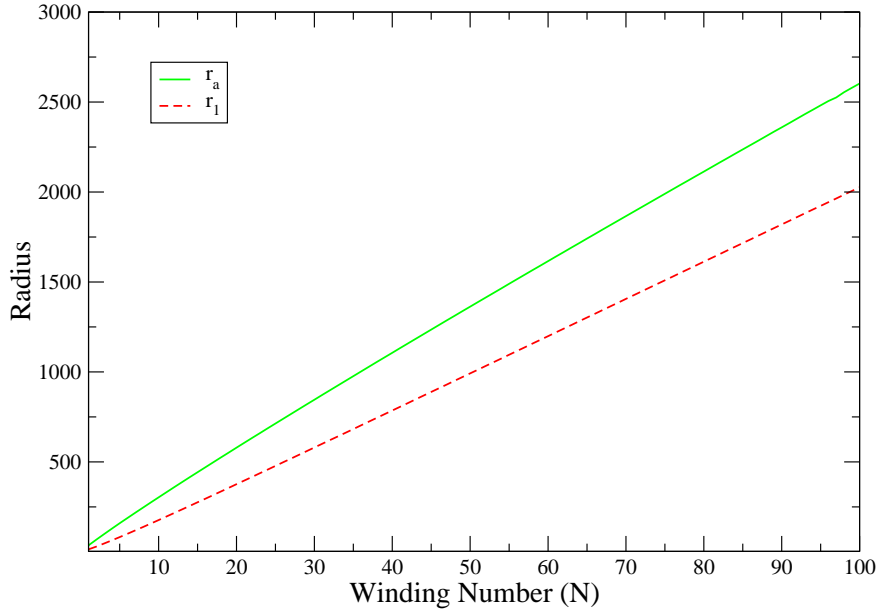}
\end{center}
\caption{Dependence of the inner scalar profile width ($r_1$) 
and vector profile width ($r_a$)
on the winding number number $N$ for a $(1,1)$ model string with
$\delta_1=\delta_2 = 1\times 10^{-20}$.
}
\label{r11nn}
\end{figure}

  In Fig.~\ref{munn} we illustrate the variation of the
tension for strings in the $(1,1)$, $(1,3)$, $(2,3)$, $(2,5)$,
and $(3,4)$ models
with the winding number $N$ for $\delta_1=\delta_2 = 1\times 10^{-20}$.
These tensions increase very slowly with $N$,
approximately logarithmically.  As the winding number increases,
the widths of the vector field profile and the inner portion of the scalar
profile do too.  This allows the angular and magnetic contributions to
the string tension to increase much more slowly than $N^2$.
The increase of the profile radii $r_1$ and $r_a$ with the winding number
$N$ is shown in Fig.~\ref{r11nn} for a $(1,1)$ model string with
$\delta_1=\delta_2=1\times 10^{-20}$.   For both $r_1$ and $r_a$,
the increase with $N$ is very close to linear.  The corresponding plots
for the other values of $(a,b)$ discussed above are nearly identical.
Unlike $r_1$ and $r_a$, varying $N$ has very little effect on $r_2$.

  We can combine the results presented above into a simple
approximate parametrization of the string tensions.
The string tension increases close to logarithmically
with the winding number $N$, but has a more complicated dependence
on $\delta_1$ and $\delta_2$, primarily through the combination
$\Delta = \delta_1+\delta_2/2$.  In the range $1<N<100$,
$10^{-30}< \Delta < 10^{-15}$, and $\delta_1$ and $\delta_2$ within
an order of magnitude from each other, the tension of a $(1,1)$ string
is reproduced to an accuracy of a few percent by the
empirical formula
\beq
\mu/\pi\,v^2 \simeq \left[\frac{4.2}{\ln(1/\Delta)}
+\frac{14}{\ln^2(1/\Delta)}\right]\;
\left(1+\left[\frac{2.6}{\ln(1/\Delta)}+\frac{57}{\ln^2(1/\Delta)}
\right]\,\ln N\right).
\label{tension}
\eea
Since the tension of an $(a,b)$ theory string is very similar to that
of a $(1,1)$ theory string for a given set of values of
$\delta_1$ and $\delta_2$, this formula also provides
a reasonable approximation to the tension
of strings in these more general theories.

  In summary, we find that the cosmic strings that arise from breaking
a $U(1)$ gauge symmetry along an almost flat direction within the $(a,b)$
models are very strongly of the Type-I variety.  The qualitative features 
of these strings can be characterized by two scales: the VEV $v$; 
and the scale of the curvature near the minimum $m$, 
which in the present case is set
by the soft supersymmetry breaking scale $m \sim \sqrt{|m_a|^2}
\sim \sqrt{|m_b|^2}\sim A$.  It is the hierarchy $m\ll v$ that makes
the potential flat.
The tension of flat-direction strings is about $\mu\sim 0.1\pi v^2$, 
while their total thickness is $w \sim m^{-1}$.
The internal structure of the strings consists of
a thin vector field core, of width close to $v^{-1}$, surrounded by
a much broader scalar profile of radius $m^{-1}$.
Flat-direction cosmic strings also have stable higher modes.
The tension of these modes grows very slowly with the winding
number $N$, increasing as $\ln N$ with a small coefficient.

  These features are much different from those of ordinary cosmic
strings derived from the abelian Higgs model, for which the relevant
scales are all on the order of the VEV $v$.  
On the other hand, the qualitative structure and the tensions of 
strings derived from the $(a,b)$ model presented above are in agreement 
with other studies of flat-direction cosmic strings~\cite{Hill:1987qx,
Freese:1995vp,Barreiro:1996dx,Perkins:1998re}.  
Within the $(a,b)$ models, we find that the form of the string profile
away from the central core and the tension can be described
well from a knowledge of  $m$ and $v$ alone, without reference 
to the precise form of the potential (or $a$ and $b$).  
This suggests that many of the results of the following sections, 
where we investigate the phenomenological features of $(a,b)$-theory 
flat-direction cosmic strings, will apply to flat-direction 
strings derived from other theories as well.


\section{String Interactions\label{interact}}

    When a pair of Type-II or weakly Type-I abelian strings 
with the same winding number intersect, 
there are effectively two possible outcomes.  
They can simply pass through each other, or they can exchange partners
and reconnect (intercommute).  When a pair of strongly Type-I $N=1$
strings collide, there is a third possibility~\cite{Bettencourt:1994kc}.
Studies of Type-I strings in the abelian Higgs model suggest that
the force between string segments is attractive.
Thus, the segments can pull together near the intersection point
to form a length of $N=2$ string, which is stable and lower
in energy than a pair of $N=1$ segments.
Under favorable conditions this segment will grow, effectively
\emph{zippering} the pair of $N=1$ strings into a single $N=2$ string.
When even higher-winding modes of strongly Type-I strings are stable as well,
we can also consider the outcome of the intersection of two strings
with general winding numbers $N_1$ and $N_2$.  Besides passing through
each other, the topology of the configuration permits the formation of
zippers with winding numbers $|N_1+N_2|$ and $|N_1-N_2|$.

  Reconnection plays an essential role in the cosmological evolution
of a cosmic string network.  It allows the network to modulate its energy
by forming string loops, which can decay away.  Without reconnection
and loop formation, the energy density in the string network could
come to dominant the universe~\cite{Hindmarsh:1994re,sbook}.
Analytic estimates and numerical simulations of Type-II and weakly
Type-I strings in the abelian Higgs model suggest that
the probability that a pair of strings
will reconnect after they intersect is close to one,
$P_r\simeq 1$~\cite{Shellard:1987bv,Matzner:1988,Achucarro:2006es}.
However, this result need not apply to very strongly Type-I strings.
These strings can form zippers, and therefore the probability
of reconnection in a string collision may differ from unity.
This can have important consequences for the evolution of a string
network in the early universe.

  In this section we investigate how flat-direction
cosmic strings interact with each other.  We begin by discussing the forces
between a pair of string segments.  Next, we study the reconnection
and zippering of strings when they intersect.  Zippering can reduce
the probability of reconnection, and it can also lead to qualitatively
new string structures that cannot be formed by Type-II strings.
We investigate how these features alter the formation of string loops.
The results of this section are applied in the sections to follow.

\subsection{Inter-String Forces}

  We found in Section~\ref{internal} above that the tension of an
$N=2$ flat-direction cosmic string is considerably lower than
twice the tension of an $N=1$ string.
Therefore bringing a pair of $N=1$ strings together (adiabatically)
from infinity to form an $N=2$ segment will lower the total energy
of the system.  As a result, we expect the (non-gravitational) force between
a pair of parallel flat-direction cosmic strings to be attractive.
More generally, we expect the interstring force to be attractive
for any other relative orientation as well.

  Our expectation is supported by both analytic estimates
of the interstring forces in the abelian Higgs model~\cite{
Taubes:1979tm,Bettencourt:1994kf,Speight:1996px},\footnote{
However, when attempting to reproduce the argument of \cite{Bettencourt:1994kf}
we found an opposite sign in the scalar term at large string
separations. Thus, we are not sure that argument is definitive.}
as well as in numerical investigations~\cite{
Jacobs:1978ch,Moriarty:1988fx,Myers:1991yh}.
It is argued in these works that the contributions to the interstring
force from the vector profile are attractive only for anti-parallel
strings and repulsive otherwise, while the scalar profile contributions
are always attractive.  For Type-I strings, the scalar profile is
wider than the vector profile and its contribution to the force has
a longer range and is always dominant.  The vector profile has a larger
range for Type-II strings explaining why the force between parallel
strings is repulsive.
The scalar profile in flat-direction strings is much wider than the
vector profile, so the results obtained in the abelian Higgs model
suggest that the force between these strongly Type-I strings
is attractive as well.

  An alternative possibility, consistent with the energetics,
is that the interstring force between flat-direction strings is
repulsive at distances larger than the string width,
and only becomes attractive when the strings overlap significantly.
Even if this were true, it would likely not have a large effect on
how these strings interact in the early universe.  Since the strings we are
studying are local (gauged), the interstring force has a very short
range, falling off exponentially outside the string core.
When a pair of strings approaches an intersection, the interstring
forces will be non-trivial only in the small region near 
the intersection point, and hence the interaction energy will be
finite.  We expect the energy required to overcome this barrier,
if it is present, to be much smaller than the initial kinetic energy
carried by the incident string segments.

\subsection{String Reconnection and Zippering}

\begin{figure}[ttt]
\begin{minipage}[t]{0.47\textwidth}
        \includegraphics[width = \textwidth]{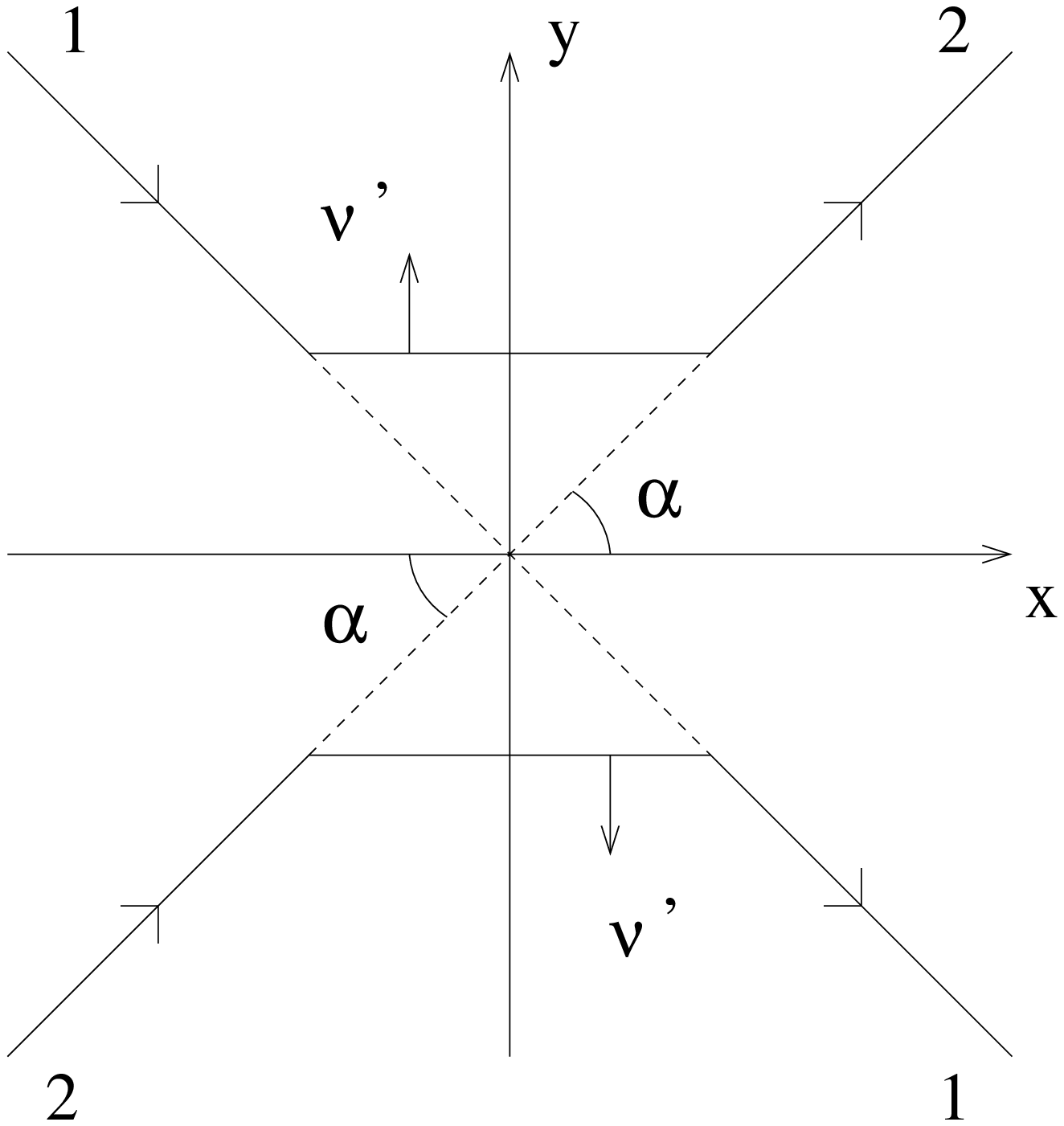}
\end{minipage}
\phantom{aa}
\begin{minipage}[t]{0.47\textwidth}
        \includegraphics[width = \textwidth]{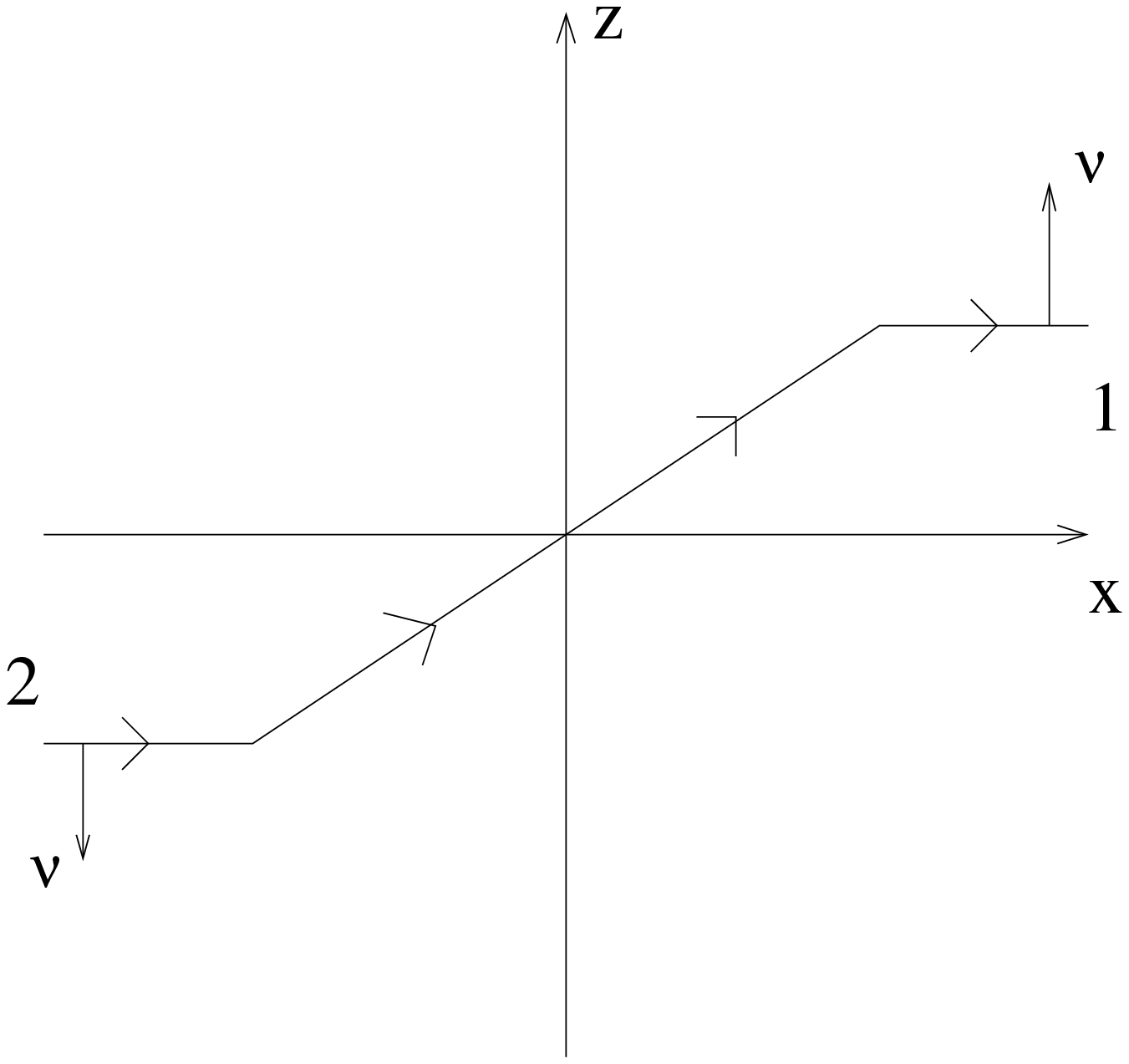}
\end{minipage}
\caption{Pictorial representation of string reconnection in
the $xy$ and $xz$ planes following Ref.~\cite{Bettencourt:1994kc}.  
The initial state consists of string $1$ and string $2$ approaching
each other along the $z$-axis, each with speed $\nu$.
In the $xz$ plane, we show only the lower string portion.
The labels $1$ and $2$ indicate which of the incident strings 
the corresponding segment was derived from.
}
\label{recxy}
\end{figure}

  A pair of strings with the same winding number
is said to reconnect (or intercommute)
if they exchange ends upon intersection.
The result of this process is illustrated in Fig.~\ref{recxy},
following Ref.~\cite{Bettencourt:1994kc}.
The initial state consists of two infinite straight strings,
each with speed $\nu$ and a relative angle $\alpha$,
approaching each other along the $z$-axis.
After exchanging ends, causality implies that 
the segments of the strings very far (spacelike-separated) 
from the intersection point continue along their 
original trajectories.  Connecting these asymptotic 
segments are new segments moving in the $\pm y$ directions.  
The labels $1$ and $2$ in the figure indicate which incident 
string the corresponding asymptotic string segment came from.  
The total length of string in the final 
configuration is clearly less than in the initial.  
Energy is conserved because the newly-formed
segments carry a velocity $\nu'$ in the $\pm y$-directions.

  Over distances that are large compared to the string width but
small compared to the horizon size, the motion of cosmic strings should
be well-described by treating them as ideal Nambu-Goto~(NG)
strings propagating in a flat spacetime background.  Therefore a
necessary condition for string reconnection is that the initial
and final configurations be kinematically allowed in the NG approximation.
It is not hard to check that for any initial relative velocity
$\nu$ and for any relative angle $\alpha$ 
(as defined in Fig.~\ref{recxy}),
this is the case~\cite{Bettencourt:1994kc}.

  The existence of a classical string solution for reconnection
does not imply that it actually occurs whenever a pair of strings intersect.
The precise outcome depends on the internal structure of the strings,
which is highly non-linear and very difficult to treat analytically.
Much of the work on this topic has therefore consisted of lattice simulations
of the corresponding classical field configurations in the abelian
Higgs model for Type-II or weakly Type-I strings.  These simulations
generally find that the probability of reconnection in a string
intersection is close to unity except for very large initial velocities,
$\nu\gtrsim 0.9$~\cite{Shellard:1987bv,Matzner:1988,Achucarro:2006es}.
Early attempts to study this question analytically,
by comparing the interaction time of the fields in the string core
to the time it takes for the pair of strings to pass through each other,
find much the same result~\cite{Copeland:1986ng}.

\begin{figure}[ttt]
\begin{minipage}[t]{0.47\textwidth}
        \includegraphics[width = \textwidth]{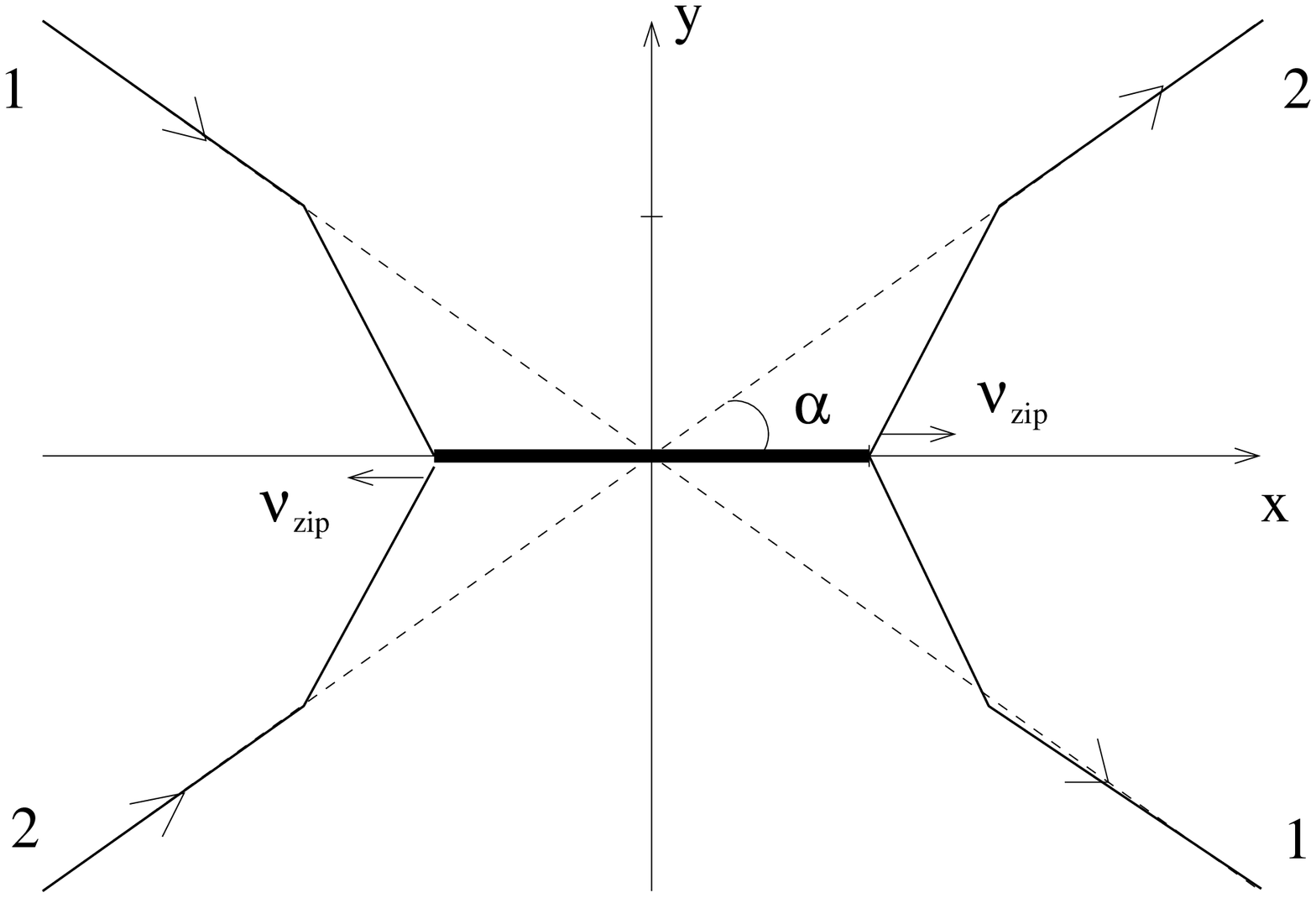}
\end{minipage}
\phantom{aa}
\begin{minipage}[t]{0.47\textwidth}
        \includegraphics[width = \textwidth]{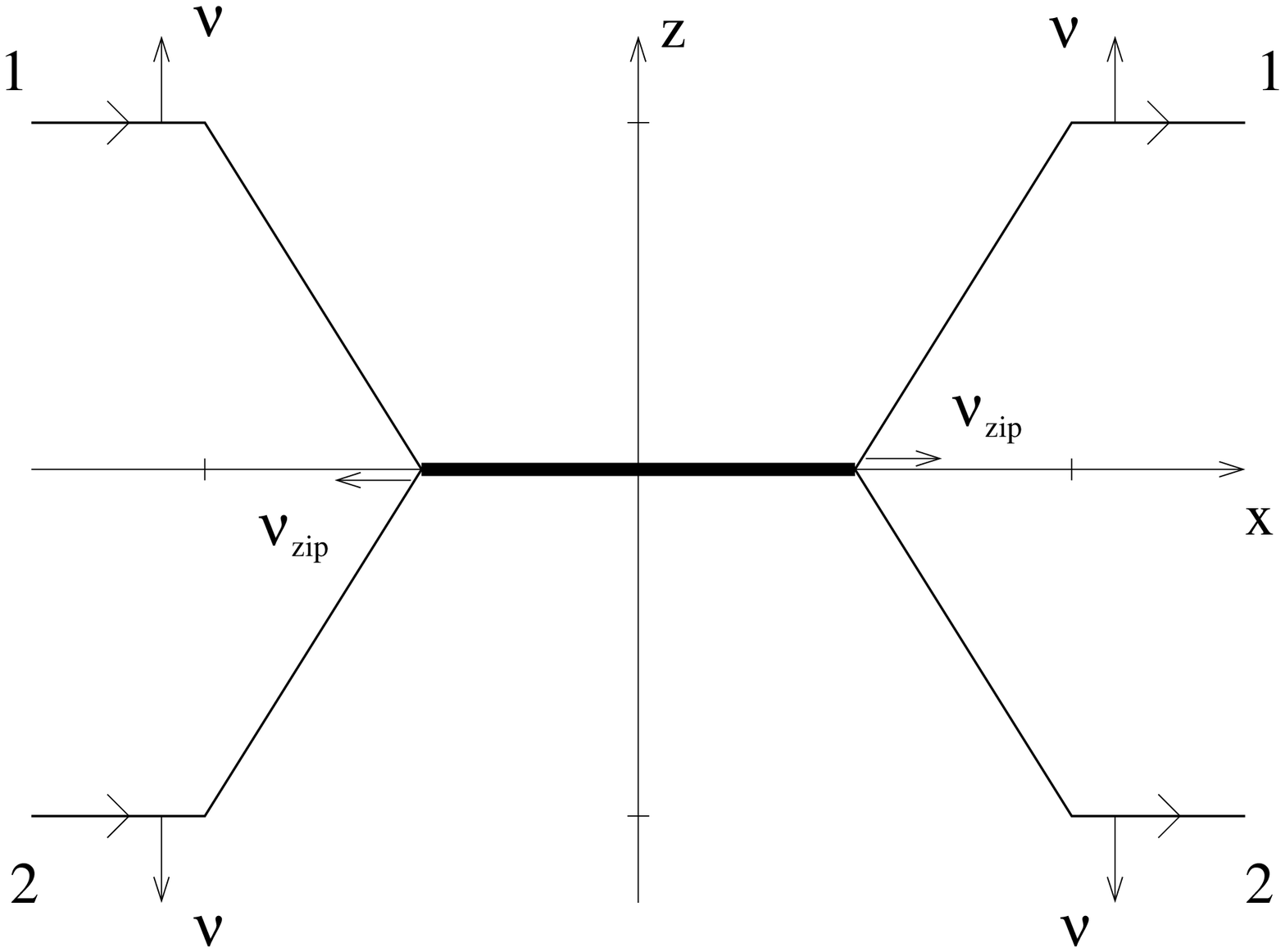}
\end{minipage}
\caption{Pictorial representation of string zippering in
the $xy$ and $xz$ planes following Ref.~\cite{Bettencourt:1994kc}.
The initial state consists of string $1$ and string $2$ approaching
each other along the $z$-axis, each with speed $\nu$.
In the $xz$ plane, we show only the lower string portion.
The labels $1$ and $2$ indicate which of the incident strings 
the corresponding segment was derived from.
}
\label{zipp}
\end{figure}

  In addition to reconnecting or simply passing through each other,
when a pair of strongly Type-I strings intersect they can also
zipper into a segment with a higher (or lower) winding number~\cite{
Laguna:1989hn,Bettencourt:1994kc,Bettencourt:1996qe}.
This is illustrated in Fig.~\ref{zipp}, 
following Ref.~\cite{Bettencourt:1994kc},
where the initial state consists of two strings with the 
same winding number $N_1=N_2=N$ approaching each other along 
the $z$-axis, each with initial speed $\nu$.
When the strings intersect, a new segment of winding number $N_{zip}=2N$ 
is formed along the $x$-axis.  This is the \emph{zipper}.  
Under favorable conditions it proceeds to grow along the $x$-axis
at the speed $\nu_{zip}$.   The string segments far from 
the intersection point (labelled by $1$ and $2$
in Fig.~\ref{zipp}) continue along their initial trajectories
on account of causality.

  Zippering has received much less attention than reconnection,
and we know of only a handful of simulations that have
studied it~\cite{Laguna:1989hn,Bettencourt:1996qe,Rajantie:2007hp}.
If string zippering is efficient, it will reduce the probability
of reconnection.  Given the importance of reconnection for the evolution
of cosmic strings in the early universe, this is a crucial issue
to be resolved.

  As for reconnection, a necessary condition for string zippering is
that it be classically allowed in the NG approximation.  Again,
this condition is only a necessary one, and the existence of a
classical zippering solution does not imply that it actually takes place.
Classical zippering solutions have been constructed in Refs.~\cite{
Bettencourt:1994kc,Copeland:2006eh}.  Unlike for reconnection,
there exist significant kinematic constraints on zippering due to
energy conservation.  For a pair of strings
with identical winding numbers $N$, initial speeds $\nu$,
and a relative angle $\alpha$, the kinematic constraint on forming
a zipper with $N_{zip} = 2\,N$ is found to be~\cite{Bettencourt:1994kc}
\beq
\sqrt{1-\nu^2}\,\cos\alpha > \frac{\mu_{2N}}{2\,\mu_N},
\label{zipconstraint1}
\eeq
where $\mu_N$ is the tension of the incident segments and $\mu_{2N}$
is the tension of the zipper.  The total length of the zippered
configuration is greater than the initial state.  Thus, a zipper can form
only if it tends to lower the energy of the configuration due to the
string tension, which requires $\mu_{2N} < 2\mu_{N}$.\footnote{
The total energy of the configuration is conserved because parts 
of the interacting string segments gain kinetic energy.}  
On the other hand, zippering does not occur
if the incident strings collide with too great a velocity $\nu$,
or if the relative opening angle between strings with the same
winding orientation is too large.

\begin{figure}[tth]
\vspace{1cm}
\begin{center}
\includegraphics[width = 0.7\textwidth]{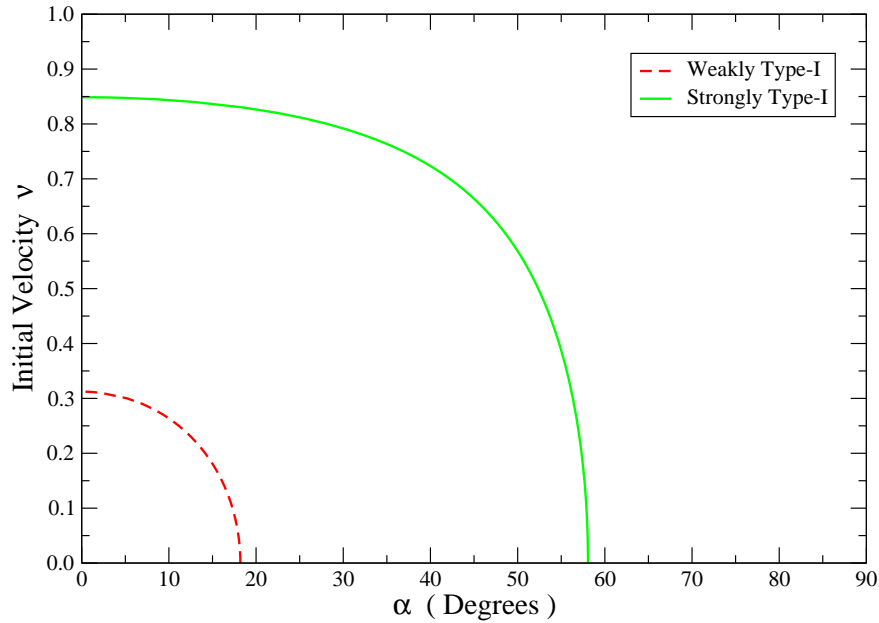}
\end{center}
\caption{
 Kinematic constraints on zippering of two $N=1$ strings to form an
$N=2$ string, in terms of the angle $\alpha$ indicated in Fig.~\ref{zipp} 
and the relative velocity $\nu$. The allowed regions lie below the curves. 
The dashed line corresponds to weakly Type-I strings, with 
$\mu_2/\mu_1 = 1.9$. The solid line corresponds to strongly Type-I 
strings associated with a flat direction potential, 
with $\Delta = 10^{-20}$, and tensions computed according to
Eq.~\eqref{tension}, which gives $\mu_2/\mu_1\simeq 1.06$.
}
\label{zip1}
\end{figure}

  In Fig.~\ref{zip1} we show the kinematic constraints on the zippering
of a pair of $N=1$ strings, in terms of the incident relative velocity
$\nu$ and the relative angle $\alpha$, defined in Fig.~\ref{zipp}.  
The region where zippering is kinematically allowed lies 
below the curves.  The dashed line for weakly Type-I strings was 
obtained assuming $\mu_2/\mu_1=1.9$.  
The solid line corresponding to the kinematic constraint on a strongly
Type-I flat-direction string was obtained using the tensions from
Eq.~\eqref{tension}, and found to be $\mu_2/\mu_1\simeq 1.06$.  
As we will discuss below, the typical relative 
velocity of a pair of strings in the early universe is expected to be less
than about $\nu\lesssim 0.7$.  Thus, zippering of flat-direction
strings in the early universe is kinematically allowed for a wide range 
of relative angles.  In the weakly Type-I case, zippering is 
only possible for small relative velocities and angles making it 
much less likely to occur.  This is why flat-direction strings can 
have a qualitatively different behavior in the early universe from 
the strings in the abelian Higgs model.  Recall that there are no 
kinematic constraints on reconnection.

\begin{figure}[ttt]
\vspace{1cm}
\begin{center}
\includegraphics[width = 0.7\textwidth]{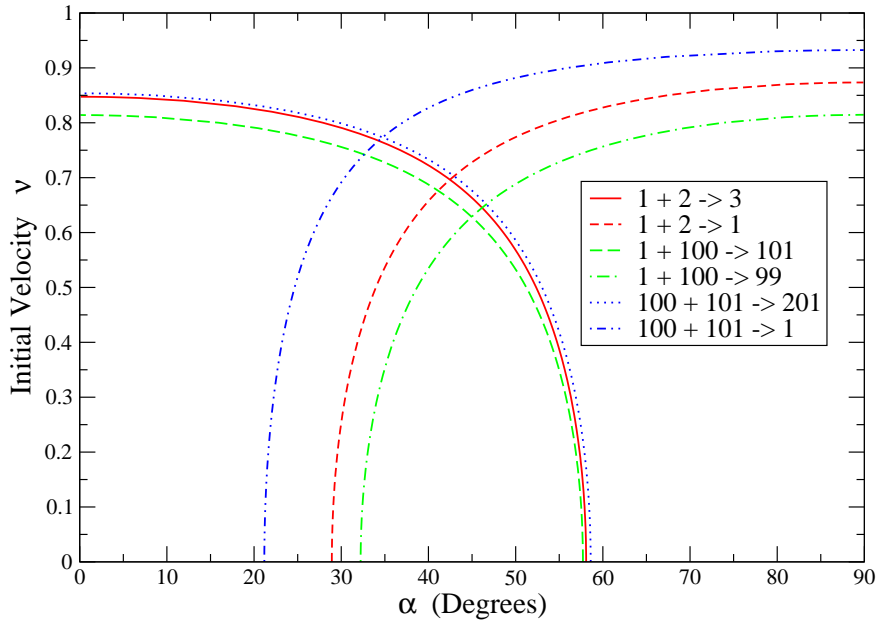}
\end{center}
\caption{
Kinematic constraints on zippering of strongly Type-I strings, for
some examples involving higher winding numbers. The allowed regions lie below
the curves. We have taken $\Delta = 10^{-20}$, and tensions computed according
to Eq.~(\ref{tension}).
}
\label{zip2}
\end{figure}

  More generally, zippering can occur between Type-I strings with
different tensions.  Incident strings with winding numbers $N_1$ and
$N_2$ can zipper into segments with $N_{zip} = (N_1 + N_2)$ or 
$N_{zip} = |N_1-N_2|$~\cite{Laguna:1989hn}.
The corresponding kinematic constraint for the zippering of strings
with unequal tensions was deduced in Ref.~\cite{Copeland:2006eh}.
Zippering is only possible when the tension of the zippered
segment is less than the sum of the tensions of the incident segments.
Even when this condition is met, zippering is only allowed
for a limited range of relative incident velocities $\nu$
and relative angles $\alpha$ (as defined in Fig.~\ref{zipp}).
We illustrate these kinematic constraints in Fig.~\ref{zip2}
for the incident string pairs $N_1 = 1$ and $N_2 = 2$, 
$N_1=1$ and $N_2=100$, and $N_1=100$ and $N_2 = 101$.
The tensions of these strings were computed using Eq.~\eqref{tension}
with $\Delta = 10^{-20}$, which applies to flat-direction strings
in the $(a,b)$ theory described in Section~\ref{internal}.
As before, the regions in which zippering is kinematically allowed
lie below the curves.  The kinematic constraints on flat-direction
strings are not overly restrictive, and zippering of various sorts
is possible over a wide range of relative velocities $\nu$
and relative angles $\alpha$ (as defined in Fig.~\ref{zipp}).

  When a pair of strings with winding numbers $N_1$ and $N_2$ intersect,
they can pass through each other, or they can form a zipper with
$N_{zip} = (N_1+ N_2)$ or $|N_1-N_2|$.  If $N_1 = N_2$, these strings can also
reconnect.\footnote{In fact, string reconnection can be treated
as the formation of a \emph{zipper} with $N_{zip} = 0$.  The classical NG
zippering solution reduces to the reconnection solution in this limit.
The absence of a kinematic constraint on reconnection can be seen
by setting $\mu_{zip} = 0$ in Eq.~\eqref{zipconstraint1}.}
There is no kinematic restriction on reconnection, and the kinematic
constraints on zippering (into one of $|N_1\pm N_2|$) are fairly mild.
Having determined the possible outcomes, it is a much more difficult
task to determine which of them actually occurs.  The answer depends on
complicated non-linear field dynamics within the string cores,
and would appear to be tractable only through numerical simulation.
Unfortunately, even this approach is further complicated 
by the large disparity in scales between the sizes of the vector 
and scalar profiles within the strings.  
Such a simulation is beyond the scope of this paper.

  To proceed, we will assume that zippering or reconnection are
likely to occur when they are kinematically allowed.  Given the high
probability of reconnection of abelian Higgs strings, this assumption
does not seem overly optimistic.
When both zippering and reconnection are possible, or when more than
one kind of zippering is allowed, we will make use of the fact that
the net force between a pair of strings is expected to be attractive.
This suggests that, near the intersection
point, the strings will pull together in whichever way is easiest.
Thus, for a pair of strings with winding numbers $N_1$ and $N_2$,
we will assume that a zipper with $N_{zip} = N_1+N_2$ forms when
$\alpha < 45^o$ (provided it is kinematically allowed), and that
$N_{zip} = |N_1-N_2|$ results for $\alpha > 45^o$.  We identify
the case $N_1-N_2 = 0$ with reconnection.

  Our assumptions are compatible with the two simulations
we know of that treat the zippering of Type-I (abelian Higgs)
strings~\cite{Laguna:1989hn,Bettencourt:1996qe}.
In both of these analyses, zippering appears to be a generic outcome
of a low-speed string intersection.  In Ref.~\cite{Bettencourt:1996qe},
the strings are found to grow until they reach the size of the box used
for the simulation, after which they pull apart.  This appears to be
the result of the boundary conditions applied to the box.
We expect that in the applications of our assumptions about string
zippering and reconnection, our qualitative results will still hold
true provided the zippering and reconnection probabilities are
of order unity.

  We end this section with a brief comment of comparison regarding
$(p,q)$ cosmic strings arising from superstring theory.
Like the flat-direction gauge-theory cosmic strings under consideration,
$(p,q)$ strings are also able to reconnect and form 
zippers~\cite{Polchinski:2004ia}.  
Even so, there are several important differences between the inter-string
interactions within these two classes of cosmic strings.  
The reconnection of $(p,q)$ strings
is a quantum mechanical process that can be related to amplitudes 
in superstring theory~\cite{Jones:2002cv,Dvali:2003zj,
Copeland:2003bj,Polchinski:2004hb,
Jackson:2004zg,Copeland:2004iv,Hanany:2005bc}.  In this sense,
it is more tractable than the non-linear classical calculation
required for field theory strings. 
It is found that the reconnection probability for $(p,q)$ strings
can be much smaller than unity, $P_r\sim 10^{-3}\!-\!1$,
depending on the underlying microscopic details.
The rules for zippering are also different for $(p,q)$ strings.
An initial state consisting of the modes $(p,q)$ and $(p',q')$
can form a zippered state with $(|p\pm p'|,q\pm q')$, which is
similar to the topological rule for Type-I field theoretic 
strings presented above.  However, a $(p,q)$ cosmic string is 
stable only if $p$ and $q$ are relatively prime integers, 
and thus the resulting zipper may sometimes decay into lower string modes.
A recent numerical simulation of a toy-model for $(p,q)$ 
cosmic superstrings has found that long-lived zippered states
are formed provided the forces between the strings are 
short-ranged~\cite{Rajantie:2007hp}.

\subsection{Loop Formation}

  Reconnection plays a crucial role in the evolution of a cosmic
string network because it is the means by which string loops form.
String loops are not topologically stable, and their decays transfer
energy out of the string network.  When cosmic strings are also able
to form zippers there are new ways for string loops to form and interact.
In the present section we enumerate some of these additional
possibilities.  We will discuss the resulting effects on the cosmological
evolution of a string network in Sections~\ref{ntwrk} and \ref{stringcosmo}.

\begin{figure}[ttt]
\begin{center}
        \includegraphics[width = 0.7\textwidth]{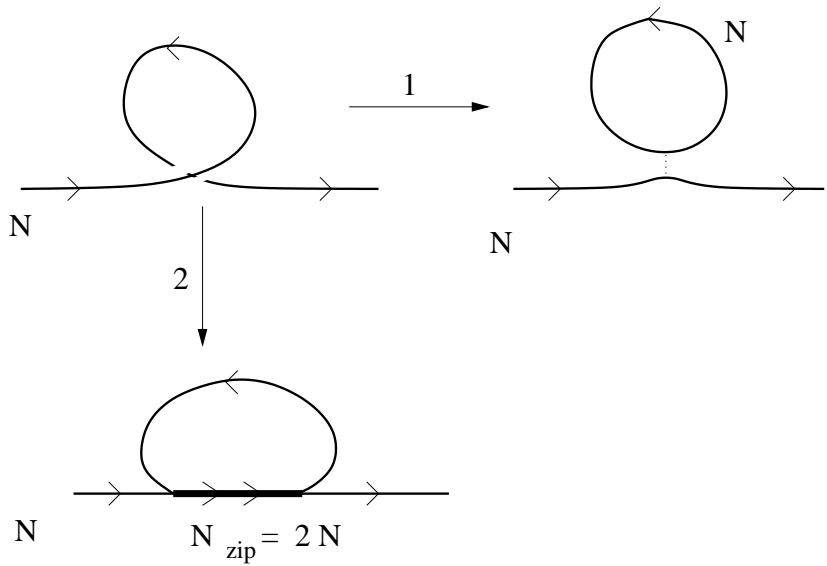}
\end{center}
\caption{Two possible ways to form a loop from the self-intersection
of a string segment.  Possibility $1$, in which a free loop is formed
by string reconnection, can occur for both Type-I and Type-II strings.
Possibility $2$, in which the loop remains connected to the parent
string by a zippered segment of a higher winding mode string, 
is only possible for Type-I strings.}
\label{selfloop}
\end{figure}

  In Fig.~\ref{selfloop} we illustrate the two ways in which
a loop can form when a string intersects itself.  The first possibility
produces a free loop through the reconnection of the intersecting segments.
This can occur for both Type-I and Type-II strings, and is the standard
mechanism for loop formation.  The loop produced
is free from the parent string.  The second possibility for loop formation
through self-intersection involves zippering of the connecting segments.
The loop formed in this way remains bound to the parent string by
a zippered segment of winding number $N_{zip} = 2\,N_1$, where $N_1$ is
the winding number of the parent.  We expect the zippered segment formed
in this way to grow until the opening angle at the junction
approaches the kinematic bound given in Eq.~\eqref{zipconstraint1}.
Subsequently, provided there are no disturbances on the string large
enough to rip the zipper apart, the bound string loop will remain attached
to the parent string as it radiates and shrinks to naught.

\begin{figure}[ttt]
\begin{center}
        \includegraphics[width = 0.7\textwidth]{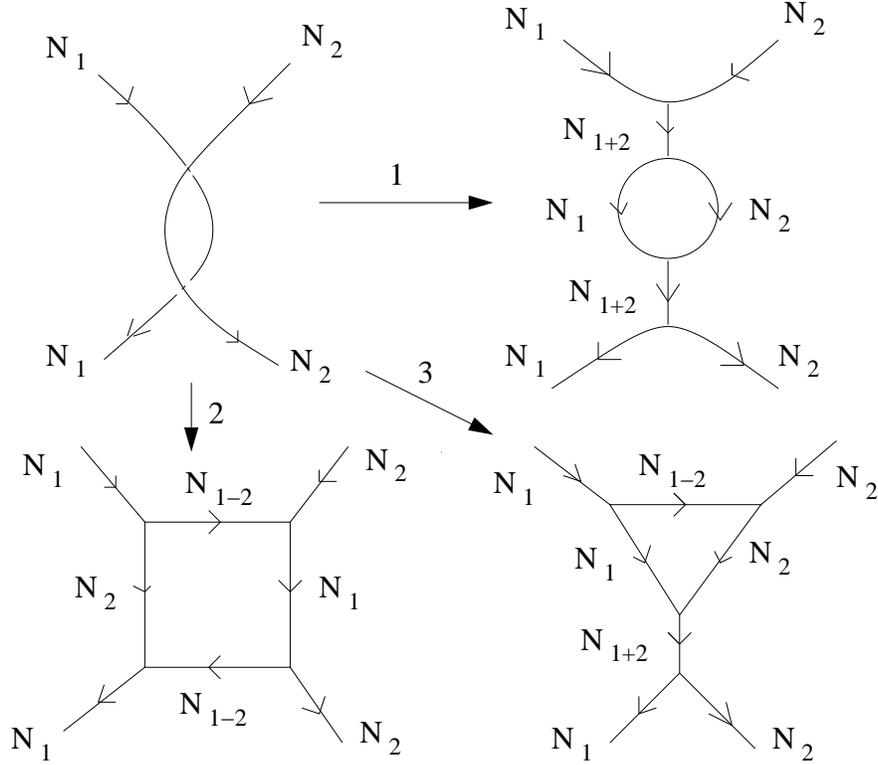}
\end{center}
\caption{Three ways to form a loop from the overlapping intersection
of a pair of Type-I cosmic strings with winding numbers $N_1$ and $N_2$.
In the figure, we have labelled the net winding number of each
string segment.  
}
\label{nonselfloop}
\end{figure}

  String loops can also be formed by the double intersection of a pair of
curved strings.  Suppose the incident strings have winding numbers
$N_1$ and $N_2$.  The topologically-allowed loops that can form
in this way are illustrated in Fig.~\ref{nonselfloop}.
At each intersection, there are two ways for the strings
to interact with each other by zippering; they can form segments
of winding number $N_{zip} = N_1+N_2$ or $N_{zip} = N_1 - N_2$.
(Here and only here, the sign of $N_i$ should be understood 
as specifying the relative orientation of the string segment.)
Possibility 1, in which both intersections produce segments of
winding $N_{zip}=N_{1+2}= N_{1}+N_{2}$ corresponds to the usual 
Type-II outcome when $N_1 = -N_2$.  Possibility $2$ has both zippered 
segments with windings $N_{zip}= N_{1-2}=N_1-N_2$.  It reduces to 
the standard Type-II case for $N_1 = N_2$.
Possibility $3$ has zippered segments with winding $N_1+N_2$ and $N_1-N_2$.
It is not immediately obvious how these configurations will evolve,
but we speculate that the loops will shrink, either through
zipper growth or loop radiation, until only a single zippered
segment remains.  The multiple outcomes shown in Fig.~\ref{nonselfloop}
also illustrate some of the many new qualitative features of
a string network consisting of strongly Type-I strings.

\section{Cosmic String Formation and Evolution\label{ntwrk}}

  Cosmic strings are much less strongly constrained by cosmology
than most other types of topological defects~\cite{Hindmarsh:1994re,sbook}.  
The reason for this is that a network of cosmic strings is 
able to regulate its energy density by forming loops, 
which radiate away.  Without loop
formation, the energy density in a cosmic string network would
scale as $a^{-2}$, redshifting more slowly than both matter ($a^{-3}$)
or radiation ($a^{-4}$), and could come to dominate the universe.
Instead, when strings are able to form unstable loops, numerical and analytic
simulations suggest that the energy density of a string network
tracks the dominant background matter or radiation density~\cite{
Bennett:1987vf,Allen:1990tv,Martins:1995tg,Vincent:1996rb,Vanchurin:2005pa}.
This behavior is called \emph{scaling}.  In the scaling regime,
the energy density of the string network makes up a fixed proportion
of about $G\mu$ of the total energy density, and this proportion
is nearly independent of the initial string density.
As long as $G\mu$ is not too large, $G\mu \lesssim 3\times 10^{-7}$~\cite{
Wyman:2005tu,Fraisse:2006xc,Pogosian:2006hg,Seljak:2006bg,Bevis:2006mj,
Battye:2006pk},
cosmic strings are generally consistent with existing cosmological bounds.

  The behavior described above was deduced from the study of
Type-II abelian Higgs string networks containing only a single
string species~\cite{Hindmarsh:1994re,sbook}. 
Strongly Type-I strings associated with supersymmetric
flat directions can modify this picture in a couple of important ways.
First, flat-direction strings have stable higher winding modes.
Even if modes with $N>1$ are not formed initially, they can be produced
as the network evolves by the zippering of lower string modes.  This opens
the possibility that flat-direction strings form a multi-tension string
network consisting of many different species.
The second reason why the evolution of flat-direction strings
in the early universe is likely to be different than for ordinary strings
is the flatness of the scalar potential.  If the $U(1)$ gauge symmetry
corresponding to the strings is restored after (or near the end of)
primordial inflation, it is likely that there will be a second,
later period of {\it thermal inflation}~\cite{Lazarides:1985ja,Lyth:1995ka}.
Flat-direction strings would be formed at the end of thermal inflation,
and hence their initial evolution is expected to be significantly
different from that of abelian Higgs strings.

\subsection{Thermal Inflation and String Formation}

  Thermal inflation occurs due to the sensitivity of 
flat potentials to thermal corrections~\cite{Lazarides:1985ja,
Yamamoto:1985rd,Lyth:1995ka}.
This flatness can be quantified by the large disparity between the
size of the curvature scale $m\sim 10^{2\!-\!3}\,\gev$ 
and the size of the VEV, $v \geq 10^{11}\,\gev$.
At the symmetry-preserving origin of the field space, there are
additional light degrees of freedom.  These induce significant corrections
to the effective potential near the origin, making it stably concave 
at high temperatures, with a curvature scale on the order 
of the temperature $T$.  For $m\ll T \ll v$, a second lower minimum
can develop far from the origin, close to the $T=0$ vacuum.
If the system begins in the symmetry preserving phase, thermal
corrections will trap it at the origin until the temperature falls down to
$T \sim m$~\cite{Yamamoto:1986jw,Barreiro:1996dx}.
While the system is trapped at the origin, it has an excess vacuum
energy on the order of $m^2v^2$.  Once the temperature of the universe
falls below $\sqrt{m\,v}$, the false vacuum energy can become dominant
and drive a period of inflation.

   Thermal inflation lasts only until $T$ falls down to $m$.
The number of $e$-foldings of expansion is therefore less than~\cite{
Lyth:1995ka}
\beq
N_i \simeq \frac{1}{2}\,\ln\left(v/m\right) \simeq 10 + \frac{1}{2}
\ln\left[\left(\frac{v}{10^{14}\gev}\right)\left(\frac{10^3\gev}{m}\right)
\right].
\eeq
This is not enough expansion to replace primordial inflation.
At the end of thermal inflation the system evolves to the
true minimum of the potential.  In this regime the constituent
fields $\Phia$ and $\Phib$ both condense, and the theory can be described
in terms of a light chiral supermultiplet corresponding to the flat
direction along with a heavy massive vector 
supermultiplet~\cite{Morrissey:2006xn}.
The scalar component of the light chiral multiplet rolls down the
potential to the true minimum and begins to oscillate.  
The false vacuum energy is transferred to 
the energy of the oscillations, which redshifts like matter, 
and dominates until the scalar field decays into radiation 
and reheats the universe.

  The reheating process can be described by the system of Boltzmann equations
\bea
\dot{\rho}_{\phi} &=& -3\,H\,\rho_{\phi} - \Gamma_\phi\,\rho_{\phi},
\label{reheating}\\
\dot{\rho}_r &=& -4\,H\,\rho_r + \Gamma_\phi\,\rho_{\phi},
\eea
where $\rho_{\phi}$ is the energy density of the scalar field oscillations,
$\rho_r$ is the energy density in radiation, $\Gamma_\phi$ is the scalar
field decay rate, and the Hubble constant $H$ is given by
\beq
H = \frac{\dot{a}}{a} = \sqrt{\frac{8\pi G}{3}\rho_{tot}}.
\label{hubbleeq}
\eeq
Here, $\rho_{tot}$ is the total energy density in the universe.
During reheating, $\rho_{tot}$ is dominated by $\rho_{r}$ and $\rho_{\phi}$.
The initial values for these evolution equations are $\rho_r \simeq m^4$,
$\rho_{\phi}\simeq m^2v^2$, and $t_i\sim 10\,H^{-1}_i\sim 10\,M_{\rm Pl}/m\,v$.
The generic value of the flat-direction decay rate is~\cite{Lyth:1995ka}
\beq
\Gamma_{\phi} = \gamma\,\frac{m^3}{v^2},
\label{decayrate}
\eeq
with $\gamma$ a constant less than or of order unity.
Once the scalars decay at about the time $t_{RH} = \Gamma_{\phi}^{-1}$,
the universe becomes radiation dominated with a reheating temperature of
\bea
T_{RH} &\simeq& 
{g_*}^{-1/4}(M_{\rm Pl}\Gamma)^{1/2}\label{trh}\\
&\simeq& 100\,\mev\;\lrf{10}{g_*}^{-1/4}\lrf{\gamma}{0.1}^{1/2}
\left(\frac{10^{14}\,\gev}{v}\right)\,
\left(\frac{m}{10^{3}\,\gev}\right)^{3/2},\nnmb
\eea
where $g_*$ is the number of relativistic degrees of freedom
at temperature $T_{RH}$ and $M_{\rm Pl} = 1/\sqrt{8\pi\,G} 
\simeq 2.4\times 10^{18}\,\gev$ is the reduced Planck mass.
The reheating temperature must exceed about $5\,\mev$ to preserve 
the predictions of nucleosynthesis~\cite{Hannestad:2004px,Ichikawa:2005vw}.
With $m = 10^{3}\,\gev$ and $\gamma = 1$, this puts an upper 
bound on $v \lesssim 10^{16}\gev$, while for $m=200\,\gev$
and $\gamma = 0.1$, the upper bound is strengthened to
$v \lesssim 10^{14}\,\gev$.  We will mostly focus on values
of the VEV less than $v\leq 10^{14}\,\gev$ for the rest of the paper.

  If flat-direction strings are to form, the corresponding $U(1)$ gauge
symmetry must be restored at or near the end of primordial inflation.
Thus, if flat-direction strings are present in the universe today,
they were most likely formed after a period of thermal inflation.
The initial densities and properties of the strings depend on the details
of the phase transition ending thermal inflation, when the flat-direction
field overcomes the thermal barrier and starts to roll down to the true
minimum.  The nature of this transition has been studied in
Refs.~\cite{Yamamoto:1986jw,Barreiro:1996dx}.
These authors find the tunnelling rate through the thermal barrier
to be negligibly small until $T\sim m$.  Below this temperature
the tunnelling suppression is not parametrically large, and bubbles
nucleate rapidly.  Of particular importance to string formation 
is the radius of the bubbles of true vacuum when they coalesce, $\xi$.  
The initial size and separation between string segments are approximately
equal to $\xi$.  Since the phase transition proceeds quickly
once the temperature falls below $m$, we expect $\xi$ to be within 
a few orders of magnitude of $m^{-1}$.

  The mechanism for string formation in the $(a,b)$ model
of flat-direction strings can be most easily understood in terms
of \emph{flux-trapping}.  The winding number of a cosmic
string is directly proportional to the net magnetic flux it carries
in its core.  In the broken phase, the magnetic flux is shielded.
As a result, random fluctuations of the gauge field in the unbroken
phase can be trapped between bubbles of broken phase.
The scalar fields surrounding tubes of trapped
flux then orient themselves to form a cosmic string with
the appropriate flux quantum number.  If $\xi$ is the typical bubble
size at coalescence, the mean winding number of the strings formed in this
way is~\cite{Rajantie:2001ps,Blanco-Pillado:2007se}
\beq
N \sim \frac{g}{2\pi}\,\sqrt{\xi\,T_f},
\eeq
where $T_f$ is the temperature at formation.
Since the phase transition proceeds quickly once $T$ falls below $m$,
we expect that $\xi$ will not be too much larger 
than $T_f^{-1}\sim m^{-1}$~\cite{Barreiro:1996dx}.  Therefore only the lowest
winding modes will be significantly populated at the beginning.
Let us also point out that the net magnetic flux of the configuration
of Eq.\,\eqref{ansatz} is $N$, independent of $a$ and $b$.

\subsection{String Network Evolution}

  Once cosmic strings are formed, their density evolves under
the influence of the spacetime expansion, as well as the processes
of reconnection and zippering.  String reconnection is particularly
important because it allows the string network to form loops and
thereby transfer its energy into radiation.  In the case of 
ordinary (abelian Higgs model) cosmic strings, the processes of 
string growth and loop production are found to balance each other, 
leading to a scaling solution.
Flat-direction strings can also interact by zippering.
This permits the formation of higher winding modes starting from an
initial population consisting only of the lowest few modes.

  Cosmic string evolution has been studied extensively through
numerical simulations~\cite{Hindmarsh:1994re,sbook,
Bennett:1987vf,Allen:1990tv}.  
However, there has been no attempt that we know of
to simulate a multi-tension string network including string zippering.
In the absence of such simulations, we turn to analytic models of string
evolution for guidance.  A number of simple models have been constructed,
and they give a good reproduction of the behavior of the long (horizon-length)
string structure seen in simulations of the abelian Higgs model.
To investigate the evolution of long flat-direction strings, 
we will make use of the model of 
Tye, Wyman, and Wasserman~(TWW)~\cite{Tye:2005fn}, which generalizes
the formulation of Ref.~\cite{Martins:1995tg}.  The TWW model was
constructed to study the behavior of long superstring cosmic strings, 
which also exhibit stable higher-winding modes and zippering, 
but with different rules for the outcome of string zippering.

 In the TWW model, long cosmic strings are characterized
by a mean velocity $\nu$, a typical correlation length along the strings
$L$, and a mean string number density $n_a$, where $a$ labels the winding
number of the string (\emph{i.e.} $N = a$).  The number density of
the string species $a$ is defined through its relation to the
energy density according to
\beq
\rho_a = \frac{\mu_a\,n_a}{\sqrt{1-\nu^2}},
\eeq
where $\mu_a$ is the tension of the species.
All string species are assumed to be described by the same $\nu$ and $L$.
This is a reasonable simplification for two reasons.
First, the tension of different strings
is a slowly varying function of the winding number, so in the absence
of interactions with other string species, each string type should evolve
in much the same way.  Second, higher winding modes are mainly formed
by the zippering of lower winding modes, and thus the speed and the
fluctuation size of different string varieties should be roughly similar.

  The evolution equations for $\nu$ and $L$ in the TWW model are taken to be
\bea
\frac{dL}{dt} &=& HL(1+\nu^2)+ {c}_1\nu,\label{leq}\\
\frac{d\nu}{dt} &=& (1-\nu^2)\left[\frac{c_2}{L}
-\nu\,\left(2H\right)\right].\label{veq}
\eea
These equations are based on the model of Ref.~\cite{Martins:1995tg},
where they are derived from the averaged equations of motion for a string
evolving in an expanding Friedmann-Robertson-Walker background 
spacetime.\footnote{Ref.~\cite{Martins:1995tg} also considers
frictional forces acting on cosmic strings.  As in the TWW model,
we do not include frictional effects in our analysis.  We have
checked that they are negligible for $v > \sqrt{m\,M_{\rm Pl}}$, 
which is expected for the flat-directions strings under consideration.}

 The TWW model generalizes  Ref.~\cite{Martins:1995tg} by adding an independent
density variable $n_a$ for each species.  The value of $n_a$ is taken to
evolve according to a Boltzmann-like equation
\bea
\dot{n}_a &=& -2\,H\,n_a -\frac{c_2\,n_a\,\nu}{L} - P_a\,n_a^2\,\nu\,L
\label{mastereq}\\
&&~~~~~+ F\,\nu\,L\,\sum_{b,c}\left[\frac{1}{2}P_{abc}\,n_bn_c\,(1+\delta_{bc})
- P_{bca}\,n_cn_a\,(1+\delta_{ac})\right].\nnmb
\eea
Here, $P_a$ is proportional to the probability of self-reconnection
for a string of variety $a$, $P_{abc}$ is the interaction probability
for the process  $b+c\to a$, and  $F$ is an overall non-self-interaction 
factor.  Once the time dependence of $H$ is specified,
Eqs.\,(\ref{leq},\, \ref{veq},\, \ref{mastereq}) form a closed 
system describing the evolution of the long string component of 
a multi-tension string network.

  The values of the constants appearing in Eqs.~(\ref{leq},\,
\ref{veq},\,\ref{mastereq}) can be fixed by comparing the scaling solution
for a single (non-interacting) string to values obtained in string simulations.
Ref.~\cite{Tye:2005fn} reports that such an agreement is obtained
with $c_1 = 0.21$, $c_2=0.18$, and $P_1= 0.28$.  We use the
same values for $c_1$ and $c_2$, which are related to the efficiency
of loop formation and the amount of small-scale structure on the strings,
respectively.  For $P_a$ and $F$, we set them to
$P_a = F = 0.28/2 = 0.14$.  Since $P_a$ is proportional to the probability
of reconnection, this accounts for our assumption that a pair of strings
is just as likely to zipper as to reconnect when both outcomes 
are kinematically allowed.  We also set the coefficients $P_{abc}$ to
\beq
P_{abc} = \left\{
\begin{array}{rl}
1;&~~~~~a = |b\pm c|,\;\:\;\nu < \nu_{\rm thresh},\\
0;&~~~~~\mbox{otherwise}.
\end{array}\right.\label{pabc}
\eeq
These values are in accord with our assumptions about zippering.
Motivated by the results of Section~\ref{interact}, we set
the velocity threshold for zippering to
$\nu_{\rm thresh}=0.85$ in our numerical analysis.

  To evaluate Eqs.\,(\ref{leq},\, \ref{veq},\, \ref{mastereq}) describing
the evolution of the string network, we must also specify the
evolution of the Hubble parameter $H$ appearing in these equations.  
We do this by solving for the scale factor $a(t)$ using Eq.~\eqref{hubbleeq}.
After thermal inflation, the two dominant sources of energy density
are $\rho_{\phi}$, from the oscillations of the light scalar field,
and $\rho_r$ for radiation.  We begin the evolution at the time
$t_i = 10\,M_{\rm Pl}/mv$, as would be expected after thermal inflation.
The initial radiation density is taken to be $\rho_r(t_i) = m^4$, while
the initial scalar field energy density is set to 
$\rho_{\phi}(t_i) = m^2\,v^2$.  After time $t_i$, $\rho_{\phi}$ and $\rho_r$
evolve according to Eq.\,\eqref{reheating}.  Since we are interested
in running the string evolution equations all the way to the present
time, we also add a very small matter density at the end of
thermal inflation, at $t_{RH}=\Gamma_{\phi}^{-1}$.
The initial matter density is chosen such that it becomes the dominant
form of energy at the approximate equality time 
$t_{eq} = 3\times 10^{36}\,\gev^{-1}$.  For reference, 
the present time is about 
$t_{0} \simeq 6.6\times 10^{41}\,\gev^{-1}$.  With $m=10^3\,\gev$, 
$v =10^{13}\,\gev$, and $\gamma = 0.1$, the initial matter density is
$\rho_m(t_{RH}) = (8.0\times 10^{-3}\,\gev)^4$.  
At later times, this dilutes according to
$d\rho_m/dt = -3\,H\,\rho_m$.
Throughout the evolution of $H$, we self-consistently assume 
that the energy density due to the string network plays 
a negligible role.

  We appeal to our expectations from thermal inflation to
set the initial values of the variables $\nu$, $L$, and $n_a$.
The symmetry breaking phase transition after thermal inflation
occurs quickly once the temperature falls below $T=m$.
The mean bubble radius $\xi$ when they coalesce
should be therefore not much larger than the nucleation radius, which
is close to $m^{-1}$~\cite{Barreiro:1996dx}.
Thus, we set $L(t_i) = 5m^{-1}$ and $n_1(t_i) = 1/(5m^{-1})^2$
as reasonable starting values.
The initial densities of the higher winding modes, $a>1$, are set to zero.
We also choose $\nu(t_i) = 0.9$.  While there is considerable arbitrariness
in these choices of initial conditions, we find that our results
at late times are largely independent of them.

\begin{figure}[ttt!]
\vspace{1.5cm}
\begin{center}
        \includegraphics[width = 0.7\textwidth]{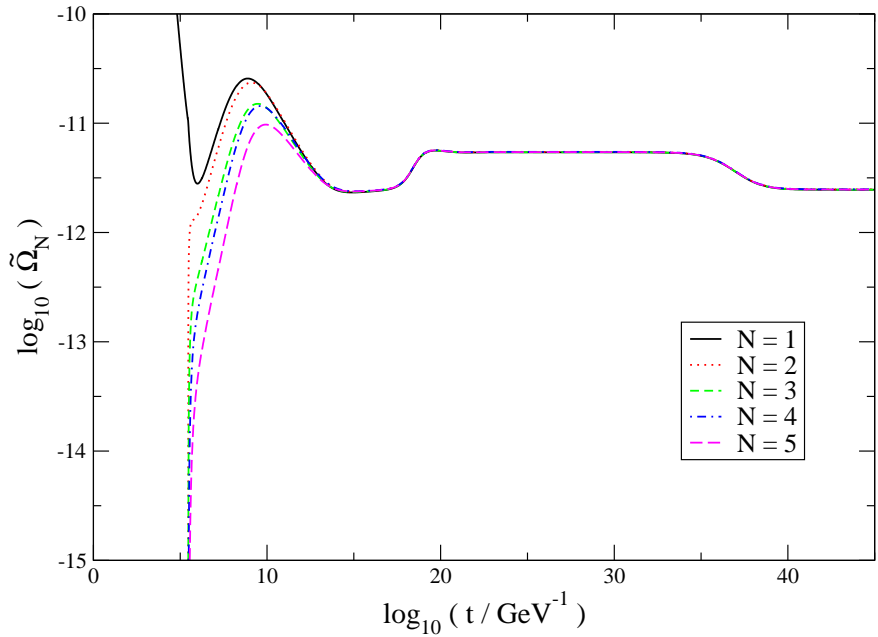}
\end{center}
\caption{Evolution of cosmic string densities
after thermal inflation with
$v = 10^{13}\,\gev$, $m=10^{3}\,\gev$, and $\gamma = 0.1$.  
We have also set $N_{max} = 50$ in generating this plot.
}
\label{tiscaling}
\end{figure}

\begin{figure}[ttt!]
\vspace{1.5cm}
\begin{center}
        \includegraphics[width = 0.7\textwidth]{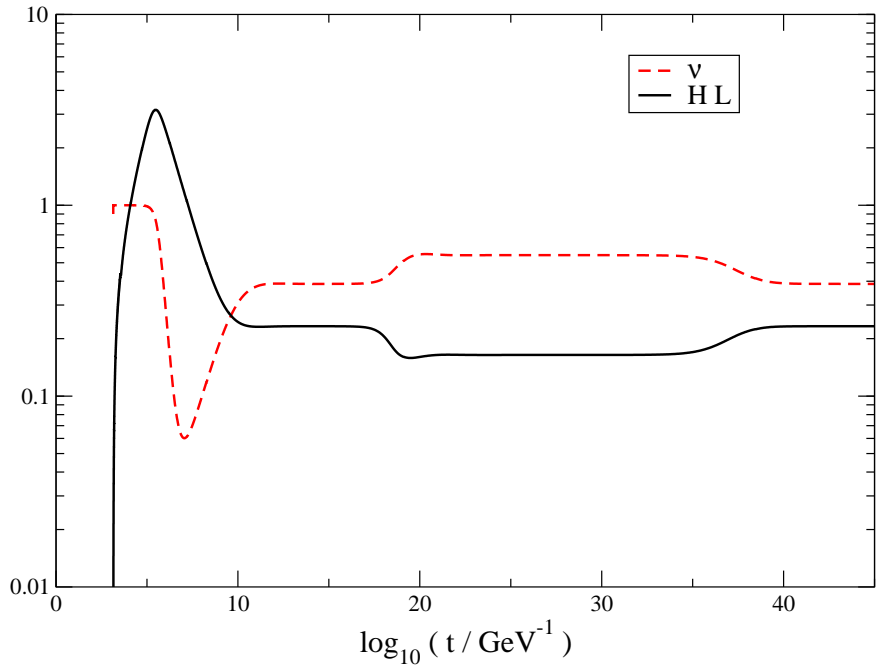}
\end{center}
\caption{Evolution of cosmic string speed and length scales
in the aftermath of thermal inflation with
$v = 10^{13}\,\gev$, $m=10^{3}\,\gev$, and $\gamma = 0.1$.  
We have also set $N_{max} = 50$ in generating this plot.
}
\label{vlscaling}
\end{figure}

  In Figs.~\ref{tiscaling} and \ref{vlscaling} 
we show the numerical solutions of the
string network equations for the model parameter values 
$m=10^3\,\gev$, $v =10^{13}\,\gev$, and $\gamma = 0.1$.  
For comparison with Section~\ref{internal}, this choice corresponds to 
a value of $\Delta = g^2m^2/v^2 \simeq 10^{-20}$.  
Fig.~\ref{tiscaling} depicts the evolution of the densities of the five
lowest winding modes in terms of the quantities
\beq
\tilde{\Omega}_a = \frac{\mu_1}{\mu_a}\Omega_a 
= \frac{\mu_1\,n_a}{\rho_c\sqrt{1-\nu^2}},
\eeq
where $\Omega_a$ is the ratio of the energy density of string species
$a$ relative to the critical density $\rho_c = 3\,H^2/8\pi\,G$,
and $\mu_a$ is the tension of string species $a$.
Normalizing by the tension makes $\tilde{\Omega}_a$ proportional 
to $n_a$ times a quantity that is independent of the winding number.
In Fig.~\ref{vlscaling} we show the evolution of the universal
length scale $L$ and universal string velocity $\nu$.

  Figs.~\ref{tiscaling} and \ref{vlscaling} show that (within the TWW model)
the string energy densities approach a \emph{scaling} solution at late times
as evidenced by $HL$, $\nu$, and $\tilde{\Omega}_a$ all flowing to constant 
values.  The scaling length, velocity, and densities are 
largely independent of the initial state of the string network.  
At late times, the string densities make up a nearly fixed 
fraction of the total energy density of the universe.  
We also find that the early era of 
oscillation dominance during reheating does not alter the final
string densities in an appreciable way.  These features are very
similar to what is found in simulations of ordinary (abelian Higgs)
string networks with only a single string species~\cite{
Bennett:1987vf,Allen:1990tv}.  

  The interesting new feature in the evolution of flat-direction 
cosmic strings is that nearly all string species flow towards 
very similar scaling values.  This is the result of string zippering, 
which allows the formation of higher winding modes from lower ones.  
Note that the formation of these higher modes does not begin immediately.  
With the initial values specified above, the initial string length scale
$L$ is much smaller than its scaling value, which is close to the
horizon scale.  This has the effect of rapidly driving the string speed
to its maximal value, $\nu\to 1$, at the outset,
as can be seen in Fig.~\ref{vlscaling},
which effectively shuts off string zippering.
Once $L$ and $\nu$ settle down to near their scaling values,
zippering begins and the higher winding-mode densities quickly
flow towards their scaling values.
This scaling behavior is quite robust.  
Changing the values of $F$ and $\nu_{\rm thresh}$
does not alter the qualitative string densities
provided $\nu_{\rm thresh}$ is larger than the mean string velocity
in the scaling regime.

\begin{figure}[ttt!]
\vspace{1.5cm}
\begin{center}
        \includegraphics[width = 0.65\textwidth]{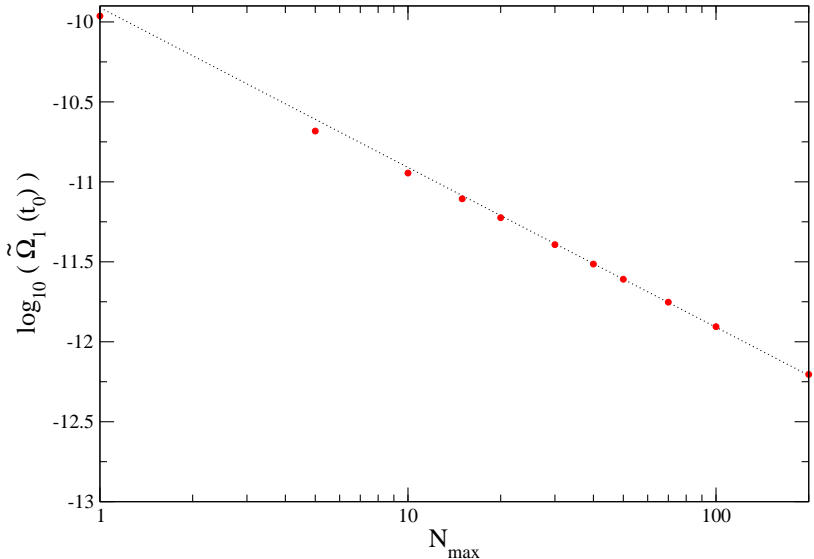}
\end{center}
\caption{Dependence of the scaling-regime string density on
the total number of string species included in the simulation, $N_{max}$.
The dotted line shows a fit to $\tilde{\Omega}_1 \propto 1/N_{max}$.
}
\label{nmax}
\end{figure}

  The fact that many string species flow towards equal scaling values
complicates the numerical analysis, since numerical limitations allow
us to include only a finite number of winding modes 
up to an unphysical maximal value $N_{max}$.
In making Figs.~\ref{tiscaling} and \ref{vlscaling} 
we have set $N_{max} = 50$.  We also find that the final, nearly universal 
scaling density of the strings depends on the artificial value of $N_{max}$.  
This feature is illustrated in Fig.~\ref{nmax}.  To a good
approximation, the near-universal string scaling density goes like
\beq
\tilde{\Omega}_a \propto \frac{1}{N_{max}},
\eeq
as illustrated by the  dotted line in Fig.~\ref{nmax}.
Evidently, the string energy density gets spread 
out among the many string types.
There is also the question of how to handle the zippering of 
strings whose winding numbers sum to greater than $N_{max}$.  
In principle, these strings can zipper into modes with $N > N_{max}$
which are not included in the simulation.  
In Figs.~\ref{tiscaling} and \ref{vlscaling} and in the  
analyses to follow, we simply disallow all such zippering processes.  
This leads to slight increase in the scaling density of modes 
with $N\gtrsim N_{max}/2$.  However, we have also studied 
other prescriptions for handling these zippering events,
and for the examples we looked at, we find qualitatively similar 
results for the modes with $N \ll N_{max}$.

  The dependence of the scaling densities on $N_{max}$ is
clearly unphysical.  We would like to take $N_{max} \to \infty$,
but this has its own problems.  Since the energy density at large $N$
goes like $\ln a$ (from the logarithmic dependence of the tension on the
winding number), if all string species flow towards a universal 
scaling density proportional to $N_{max}$
the total network energy density goes like
\beq
\rho_{tot} \propto \frac{1}{N_{max}}\sum_{a=1}^{N_{max}}\,
\ln a \simeq \ln N_{max}.
\eeq
This diverges logarithmically as $N_{max}\to \infty$.
In practice, however, this divergence is not realized.
The initial string spectrum consists almost entirely
of the lowest modes, the density of higher modes is built
up from the lower modes by zippering, and these higher modes take
longer to reach their scaling values. At any given time,
only a finite number of strings have developed their 
scaling density.\footnote{In this sense, our use of the term 
\emph{scaling} for flat strings is somewhat 
more general than its meaning for ordinary cosmic strings because 
the string densities are not completely static, but very slowly varying.}
Let us define $N_{eq}(t)$ as the highest mode that has reached scaling 
by time $t$.  Modes with $N > N_{eq}(t)$ all have densities well below
their equilibrium scaling values.  Thus, at time $t$,
we effectively have $N_{max} = N_{eq}(t)$, and the total
energy contained in the string network goes like $\ln N_{eq}(t)$.

\begin{figure}[ttt!]
\vspace{1.5cm}
\begin{center}
        \includegraphics[width = 0.75\textwidth]{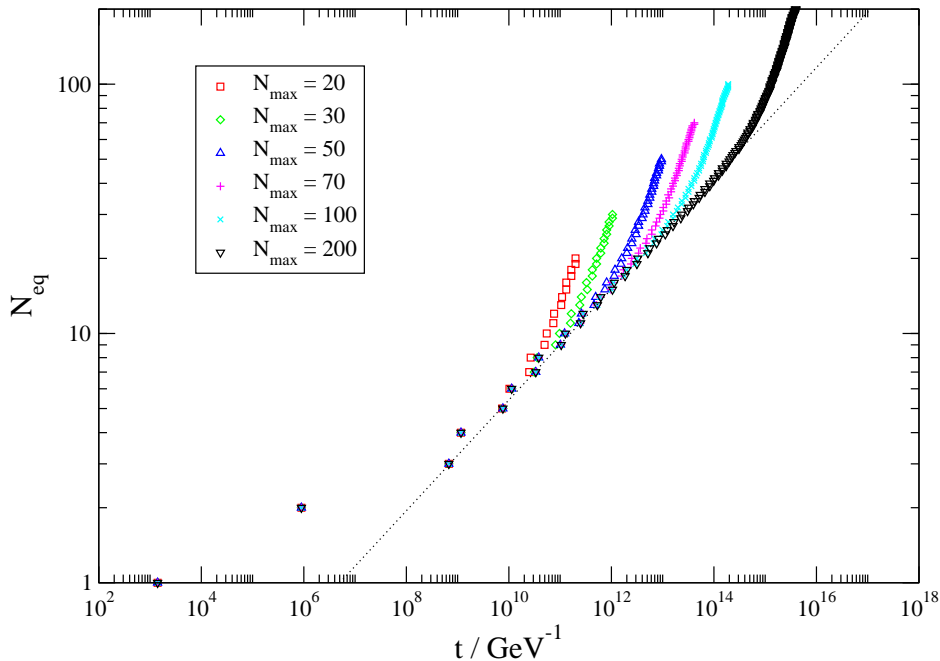}
\end{center}
\caption{The number of string species that have reached scaling, $N_{eq}$,
as a function of time in the aftermath of thermal inflation 
for different values of $N_{max}$, with the parameter values 
$v = 10^{13}\,\gev$ and $m=10^{3}\,\gev$.
The dotted line indicates an approximate fit to 
$N_{eq}(t) \propto t^{0.22}$ in the region where 
the curves appear to be universal.}
\label{neqscaling}
\end{figure}

  In Fig.~\ref{neqscaling} we show the time evolution of $N_{eq}(t)$
for several values of $N_{max}$.  All other parameters are the same
as in Figs.~\ref{tiscaling}.
The curves for different values of $N_{max}$ match up for 
$N \lesssim N_{max}/3$, but start to deviate from each other 
as the winding number $N$ approaches $N_{max}$.  
Focusing on the apparently universal portion of these curves, 
the rate of increase of $N_{eq}(t)$ with time goes like $t^{0.22}$.  
If we can extrapolate this dependence to much larger winding numbers, 
the value of $N_{eq}$ at the present time $t_0$ will be
\beq
N_{eq}(t_0) < \lrf{t_0}{t_i}^{0.22} \simeq 10^{8},
\eeq
where we have used $t_i \simeq 10\,M_{\rm Pl}/{m\,v} \simeq 2.4\times 
10^3\gev^{-1}$ and $t_{0}\simeq 6.6\times 10^{41}\,\gev^{-1}$.
This is a very large number, but it is not so large so as to be problematic.
Recall that the string tension, given in Eq.~\eqref{tension}, increases
logarithmically with the winding number.  The tension of
a string with $N = 10^{8}$ is merely
\beq
\mu_{N} < 3\,\mu_1,
\eeq
for $m = 10^3\,\gev$ and $v = 10^{13}\,\gev$, corresponding to
$\Delta \simeq 10^{-20}$ in Eq.~\eqref{tension}.
Moreover, the total string energy density in the network 
is less than about $\ln N_{eq}(t_0) \lesssim 20$ times the
energy density of a network containing a single type of string
with the same tension as the lowest mode.
These values for the maximal tension and the total string
density are not much larger than for an ordinary cosmic string,
and they present no obvious cosmological difficulties.

  In our analysis of flat-direction string network evolution described above 
we have used a very simple analytic model of string network evolution;
we have made specific assumptions about the details of the string interactions;
and, we have made extrapolations into regions well beyond what we are 
able to probe analytically and numerically.  
Nevertheless,  a simple picture for the evolution
of a flat-direction string network emerges from our results, 
and is likely to be genuine,
even if some of the underlying assumptions are not necessarily rigorous
and the model used to study the network evolution is overly simple.
In this picture, a very large number of string species reach
similar scaling densities by the present time.  The total energy of the
network is within an order of magnitude or two of the energy density
that a single abelian Higgs string species would have for the same value of 
the string tension.  However, instead of being concentrated within 
a single species, the string energy density is nearly uniformly 
distributed among all the string species that have attained scaling.  
Thus, the flat-direction string network consists of a near 
\emph{continuum} of string species, but with global properties
that closely resemble those of a single species network.

  Our argument for this picture is based on the very slow dependence
of the flat-direction string tensions on the winding number.
On account of this slow variation, the macroscopic properties of
the many species that have attained scaling are very similar to each other.
For example, Fig.~\ref{munn} shows that the tension of a mode with
$N=100$ is only about $1.4$ times that of the $N=1$ mode for
$\Delta \simeq 10^{-20}$.  From this feature, as long as the zippering is
reasonably efficient and the lowest mode is able to attain a
scaling value for its density, we expect the densities of 
the string modes to be very similar to one another up to
large values of the winding number $N\gg 1$.  

  One curious aspect of this picture is that the total energy density
in the network corresponds to less than a few hundred individual
strings of horizon length.  It is therefore curious that the TWW
model applied to flat-direction strings predicts that there are many 
more string species than this in the scaling regime at the present time, 
each with a characteristic length scale of horizon size.  
We  suggest that the scaling densities predicted by 
the TWW model for flat-direction strings should be interpreted as 
time-averaged values.  At any given epoch in the scaling regime, 
there exist many fewer long strings than $N_{eq}(t)$.  
However, these strings are continually zippering into other string species, 
and averaged over time, many more string species are populated 
(with a lower density per string) than are present at any one time.  
It is also possible that this issue of discreteness leads to a value of
$N_{eq}(t)$ that is smaller than what is predicted by the TWW model.

  A definite confirmation of this picture of flat-direction string evolution
would appear to require a full numerical simulation of the network 
(as well as lattice simulations to determine the zippering probabilities).  
This task is complicated by the need to include many different string species
in the simulation, and is beyond the scope of this paper.
We have, however, examined the effect of changing some of our
assumptions about string zippering encoded in the coefficients
$P_{abc}$, defined in Eq.~\eqref{pabc}.  For example, we find that
reducing the probability for zippering into $a = (b+c)$ relative
to $a=|b-c|$ does not significantly alter the final scaling densities.
We have also looked into modifying the interaction terms in 
Eq.\,\eqref{mastereq}, as suggested in Ref.~\cite{Avgoustidis:2007aa}, 
and we again find the same qualitative picture of string network evolution.
These results suggest that the picture of
flat string evolution presented here is robust.

  Before moving on, let us briefly compare our near-\emph{continuum} 
picture of flat-direction cosmic strings to the 
cosmological picture of $(p,q)$ surperstring cosmic strings 
derived in Ref.~\cite{Tye:2005fn}.
These quasi-fundamental strings can be labelled by
pairs of integers $(p,q)$ with $p\geq 0$.  A string state is stable
only if $p$ and $q$ are relatively prime.  States with $(p,q)$ not
relatively prime can be formed but are only marginally stable.
They are expected to decay into lower, stable modes after they are created.  
In the analysis of Ref.~\cite{Tye:2005fn}, this additional dissipative
channel led to a rapid decrease in the relative population 
of higher-tension modes.
That superstring cosmic strings do not form a near-continuum scaling
network is also not surprising given that the tensions of these strings
increase fairly rapidly with the mode numbers~\cite{Tye:2005fn},
\beq
\mu_{(p,q)} \propto \sqrt{g_s^2\,p^2+q^2},
\eeq
where $g_s$ is the superstring coupling.\footnote{
This formula applies in ten-dimensional flat space.
It may receive corrections in other backgrounds~\cite{Firouzjahi:2006vp}.}
Hence, even though flat-direction cosmic strings and $(p,q)$ strings can both
form stable winding modes through zippering, these two varieties of cosmic
strings interact and evolve in significantly different ways.


\section{String Signatures\label{stringcosmo}}

  If cosmic strings are present in the early universe
they can give rise to a number of observable signatures.
No evidence for cosmic strings has been found in the temperature
power spectrum of the cosmic microwave background or in large-scale 
sky surveys.  This implies the constraint $G\mu \lesssim 3\times 10^{-7}$,
fairly independently of the underlying string model~\cite{
Wyman:2005tu,Fraisse:2006xc,Pogosian:2006hg,Seljak:2006bg,Bevis:2006mj}.
Beyond these limits,
the most promising signatures for ordinary (abelian Higgs) 
cosmic strings are gravitational lensing and gravitational 
radiation~\cite{Hindmarsh:1994re,sbook}.
We find that these signals can be modified for flat-direction
cosmic strings.  Flat-direction cosmic strings are also more
likely to radiate into their constituent particles than ordinary
cosmic strings, leading to new classes of potential signatures.  
By combining observations of several different phenomena,
it may be possible to distinguish flat-direction cosmic strings
from ordinary cosmic strings as well as $(p,q)$ cosmic superstrings.

\subsection{Gravitational and Particle Radiation from Loops}

  Cosmic strings emit gravitational radiation primarily through
the oscillations of string loops.  For both ordinary and flat-direction
cosmic strings, a single loop is expected to emit
gravitational radiation with power
\beq
P_{gw} = \Gamma\,G\mu^2,
\label{pgw}
\eeq
where $\Gamma=10\!-\!100$ is a dimensionless constant whose precise value
depends on how the loop is oscillating~\cite{Vilenkin:1981bx,Turok:1984cn,
Burden:1985md,Vachaspati:1984gt,Garfinkle:1987yw}.  
This rate is independent of the length of the loop, $\ell$.
The radiation frequencies do depend on $\ell$ and are
\beq
f_n = \frac{2\,n}{\ell},~~~~~n=1,2,3,\ldots
\label{omega}
\eeq
with the relative power going into mode $n$ decreasing at least 
as quickly as $n^{-4/3}$ for simple string loop 
solutions~\cite{Turok:1984cn,Burden:1985md,Vachaspati:1984gt,Garfinkle:1987yw}.

  To compute the gravitational wave background from a cosmic string
network, one must convolute the power emitted by individual loops with
the loop density distribution.  Unfortunately, even for ordinary cosmic
strings, the loop density distribution is not fully understood.
The main uncertainty is the size of loops when they are formed.
It is standard to parametrize the typical initial loop length according to
\beq
\ell_i = \alpha\,t,
\eeq
where $t$ is the time of loop formation, and 
estimates for $\alpha$ range between the 
string width~\cite{Vincent:1996rb}, 
to $(\Gamma\,G\mu)^{\chi}$ with $\chi \geq 1$~\cite{Siemens:2002dj,
Polchinski:2006ee,Polchinski:2007qc}, 
all the way up to $\alpha = 0.1$~\cite{Vanchurin:2005yb}.
We will consider different values of $\alpha$ below.

  Cosmic string loops can also radiate directly  
into particles~\cite{Srednicki:1986xg,Damour:1996pv,Peloso:2002rx}.
This can arise both through the direct emission of particles from 
smooth strings~\cite{Srednicki:1986xg,Damour:1996pv,Peloso:2002rx}, 
as well as from \emph{cusp annihilation}~\cite{Brandenberger:1986vj}.  
For the string loops present in the early universe, 
cusp annihilation is usually the more important source of 
particle emission~\cite{Brandenberger:1986vj}.\footnote{
This conclusion can change if there exist light (superstring) 
moduli fields with masses much smaller than $w^{-1}$, 
where $w$ is the width of the string~\cite{Damour:1996pv,Peloso:2002rx}.
For flat direction strings, both the string width and the typical
moduli mass are set by the scale of supersymmetry breaking $m$.
As a result, the rate of moduli emission by flat-direction strings
is suppressed, and the corresponding bounds~\cite{Babichev:2005qd} 
are not relevant.
}
A \emph{cusp} is a point on a string that reaches the speed of
light at some instant during its (Nambu-Goto) evolution.  
Cusps are a generic feature of many simple solutions 
for the motion of a string loop, where they are found 
to occur about once per oscillation period~\cite{Turok:1984cn,sbook}.  
In the region near the cusp, the string segments fold back 
upon themselves such that the separation between 
the adjacent segments becomes smaller than the string width.
This allows these string segments to annihilate each other.
Cusps should not be confused with string \emph{kinks}, which
are points on a string where the tangent vector changes substantially
over a very short distance, on the order of the 
string width~\cite{Garfinkle:1987yw}.
Unlike at a cusp, there need not be any significant annihilation
of the string segments in the vicinity of a kink, 
and kinks can persist for many loop oscillations~\cite{Quashnock:1990wv}.
Kinks can be created from string reconnection and zippering.

  The effective length of the overlap region between the adjacent
string segments near a string cusp on a loop of length $\ell$ is about
\beq
\ell_c = \sqrt{w\,\ell},
\eeq
where $w$ is the string width~\cite{BlancoPillado:1998bv}.
The overlapping string segments near the cusp are expected
to annihilate, transferring most of the string energy within the
overlap region to the constituent particles making up the string.
The total average power released into particles through this process 
by a single string loop is~\cite{BlancoPillado:1998bv}
\beq
P_{cusp} \simeq {\mu}\,\ell_c\,\lrf{c}{\ell}.
\eeq
Here, $c/\ell$ is the cusp rate, where $1/\ell$ corresponds to
the period of a loop oscillation, and $c$ is the probability 
per period for a cusp to occur.
We expect $c\sim 1$, although it has been argued that the presence 
of \emph{kinks} on strings could push it to smaller 
values~\cite{Garfinkle:1987yw}.  This is yet another uncertainty 
associated with the structure of cosmic strings on small scales.

  For a given tension, flat-direction strings are much wider
than ordinary cosmic strings; $w\sim m^{-1} \gg v^{-1}$ 
compared to $w\sim v^{-1} \sim \mu^{-1/2}$.
The amount of string annihilated in a cusp is therefore 
greatly enhanced.  The total particle radiation power from 
cusp annihilation by a flat-direction string loop is 
\beq
P_{cusp} \simeq \frac{c\,\mu}{\sqrt{m\,\ell}}.
\eeq
Relative to the gravitational radiation power, Eq.~\eqref{pgw}, 
we see that cusp annihilation dominates for sufficiently small loop sizes.  
The loop size at which the two powers become equal is
\beq
\ell_{=} \simeq m^{-1}\lrf{c}{\Gamma G\mu}^2.
\eeq
Recall that the loop size at formation is $\ell_i = \alpha\,t$.
For $\ell_i \lesssim \ell_=$ the loops will decay primarily through
particle emission, and not gravitational radiation.  
On the other hand, when $\ell_i \gg \ell_=$, 
most of the loop energy will go into gravity waves, except
for a small burst of particles towards the end of the loop's existence.

  Thus, the emission of particles by flat-direction cosmic 
strings through cusp annihilation
is greatly enhanced relative to ordinary cosmic strings.
If cusp annihilation dominates over gravitational radiation,
many of the gravitational radiation signals will be suppressed 
compared to ordinary cosmic strings.
In order to compare the relative signals from gravitational radiation
and particle emission, it is helpful to concentrate on three particular
epochs in the early universe:
the reheating time $t_{RH}$; the time at which $\ell_i=\ell_= = \alpha\,t_=$; 
and the earliest time $t_f$ at which a given gravitational wave frequency
mode $f$ can form.  

  We found in Section~\ref{ntwrk} that reheating after thermal
inflation occurs when $t \simeq t_{RH} := \Gamma_{\phi}^{-1}$, where 
$\Gamma_{\phi} = \gamma\,m^3/v^2$ is the decay rate of the light
flat-direction scalar field.  This yields
\beq
t_{RH} = \lrf{0.1}{\gamma}\lrf{v}{10^{14}\,\gev}^2\lrf{10^3\,\gev}{m}^3
(10^{20}\,\gev^{-1}).\label{timerh}
\eeq
Recall that if the process of reheating after thermal inflation
is to avoid disturbing the predictions of nucleosynthesis,
we must have $v\lesssim 10^{16}\,\gev$ for $m=1000\,\gev$ and $\gamma = 1$,
and $v\lesssim 10^{14}\,\gev$ for $m=200\,\gev$ and $\gamma = 0.1$.  

  The second moment of interest, the time after which 
newly-formed loops lose most of their energy in the form 
of gravity waves, occurs when $\ell_i=\ell_= =\alpha\,t_=$.  
This corresponds to the time
\bea
t_= &=& \alpha^{-1}\lrf{c}{\Gamma G\mu}^2m^{-1}\label{tee}\\ 
&\simeq&\alpha^{-1}\,c^2\,\lrf{50}{\Gamma}^2\lrf{2\times 10^{-11}}{G\mu}^2
\lrf{10^3\,\gev}{m}\,(10^{15}\,\gev^{-1}).\nnmb
\eea
We have expressed $t_=$ in terms of $G\mu$ rather than the VEV $v$
because it is this dimensionless combination that appears frequently
in the estimates below.  An approximate conversion between 
$G\mu$ and $v$ is (see Fig.~\ref{mud2})
\beq
G\mu \simeq \lrf{v}{10^{14}\,\gev}^2\,(2\times 10^{-11}).
\eeq
Given the upper bound on $v$ from reheating after thermal inflation,
we will mostly focus on $v\lesssim 10^{14}\,\gev$.

  The third time of interest is $t_f$, the earliest moment at which 
a given gravitational wave frequency as low as $f$ can be emitted.  
Recall that loops
formed at time $t_i$ have the initial size $\ell(t_i)=\alpha\,t_i$, 
and subsequently shrink and radiate into frequencies $f\geq 2/\ell$.  
For a mode observed at the present time with frequency 
$f = f(t_0)$ emitted at time $\tilde{t}$, the initial frequency was
\beq
f(\ttt) = \frac{a(t_0)}{a(\ttt)}\,f.
\eeq
Combining these facts, the earliest time $t_f$ at which a mode with present
frequency $f$ could have been emitted is
\bea
t_f &=& \frac{2}{\alpha\,f}\frac{a(t_f)}{a(t_0)}\label{tff}\\
&\simeq& \left\{
\begin{array}{lcc}
\alpha^{-3}\lrf{10^{-7}Hz}{f}^3\,(6.5\times 10^{10}\,\gev^{-1})
&~&t_f>t_{eq}\\
\alpha^{-2}\lrf{10^{-7}Hz}{f}^2\,(2.5\times 10^{19}\,\gev^{-1})&~&
t_{RH}<t_f<t_{eq}\\
\alpha^{-3}\lrf{10^{-7}Hz}{f}^3\,\lrf{10^{14}\,\gev^{-1}}{t_{RH}}^{1/2}
(1.2\times 10^{22}\,\gev^{-1})&~&t_f<t_{RH}
\end{array}\right..\nnmb
\eea

  Both $t_=$ and $t_f$ depend on the parameter $\alpha$ that characterizes
the typical size of a string loop when it is formed, $\ell_i = \alpha t_i$.
The dynamics of loop formation are not completely understood,
and as a result, estimates for $\alpha$ vary widely.  
Some recent simulations find that a significant portion of the 
loops formed are quite large, with $\alpha \simeq 0.001$~\cite{
Ringeval:2005kr,Martins:2005es} or $\alpha \simeq 0.1$~\cite{Vanchurin:2005yb}.
Other simulations find that the typical initial loop size 
approaches their resolution 
limits~\cite{Vincent:1996rb}.  In this case, it is thought that 
gravitational radiation will smooth out very small
fluctuations, and impose a lower limit on $\alpha$~\cite{sbook}.  
The scale over which this smoothing occurs is also under ongoing
investigation.  Early estimates suggested $\alpha = \Gamma\,G\mu$~\cite{sbook},
but more recent analyses have found even smaller values of $\alpha$.
In Ref.~\cite{Siemens:2002dj} the authors obtain 
$\alpha = (\Gamma\,G\mu)^{\chi}$
with $\chi=1.5$ during the radiation era and $\chi=2.5$ during 
matter dominance.  The authors of Ref.~\cite{Polchinski:2006ee} 
find $\alpha \simeq 0.6\,\Gamma\,(G\mu)^{\chi}$ with $\chi = 1.2$ 
in the radiation era and $\chi=1.5$ in the matter era.
Furthermore, in Ref.~\cite{Polchinski:2007qc} 
it is suggested that the simulation results of Ref.~\cite{Vanchurin:2005yb}
should be interpreted as predicting a network
with $10$-$20\%$ of the loop energy density in the form of large loops
with $\alpha \simeq 0.1$ and  the remainder in the form of very small loops
with $\alpha \simeq \Gamma(G\mu)^{\chi}$ with $\chi>1$.
On account of the rapidly evolving state of the field, we will
consider both large and small values of $\alpha$ below.

\begin{figure}[ttt!]
\vspace{1.5cm}
\begin{center}
        \includegraphics[width = 0.75\textwidth]{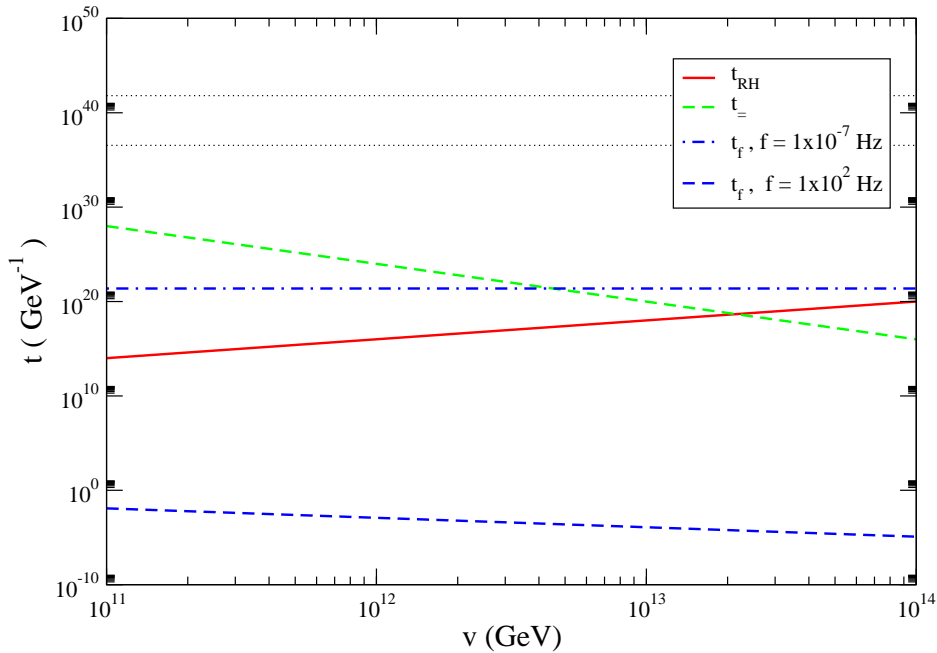}
\end{center}
\caption{Dependence of the times of interest $t_{RH}$, $t_=$,
and $t_f$ on the VEV $v$ for $\alpha = 0.1$.  The black dotted lines 
indicate the present time $t_0 \simeq 6.6\times 10^{41}\,\gev^{-1}$
and the matter-radiation equality time 
$t_{eq}\simeq 3.5\times 10^{36}\,\gev^{-1}$.
}
\label{times-big}
\end{figure}

  Fig.~\ref{times-big} shows $t_{RH}$, $t_=$, 
and $t_{f}$ in $\gev^{-1}$ units as functions of the VEV $v$
for large loops with $\alpha = 0.1$.  
The model parameters were set to $m=10^3\,\gev$, $c=1$, 
$\gamma = 0.1$, $\Gamma = 50$.  In this plot we also indicate 
the present time $t_0 \simeq 6.6\times 10^{41}\,\gev$ and the 
matter-radiation equality time $t_{eq} \simeq
3.5\times 10^{36}\,\gev$ with dotted lines.
The value of $t_f$ is shown for two values of the frequency,
$f=10^{-7}\,Hz$ and $f=10^2\,Hz$.  These values span most of
the range relevant for gravitational wave searches.  
For these large loops, $t_=$ is always much less than $t_{eq}$, and
all loops formed after $t_=$ will decay predominantly into 
gravitational radiation.  At the lower frequency $f=10^{-7}\,Hz$, 
$t_f$ lies below $t_{eq}$ but above $t_{RH}$, and is never much 
less than $t_=$.  This suggests that the gravitational wave signal 
at this frequency will not be attenuated much by the enhanced
rate of particle emission by the loops.  On the other hand,
the value of $t_f$ for $f=10^2\,Hz$ lies well below both $t_{RH}$
and $t_=$, indicating that the high-frequency gravitational wave
signal will be reduced.  

\begin{figure}[ttt!]
\vspace{1.5cm}
\begin{center}
        \includegraphics[width = 0.75\textwidth]{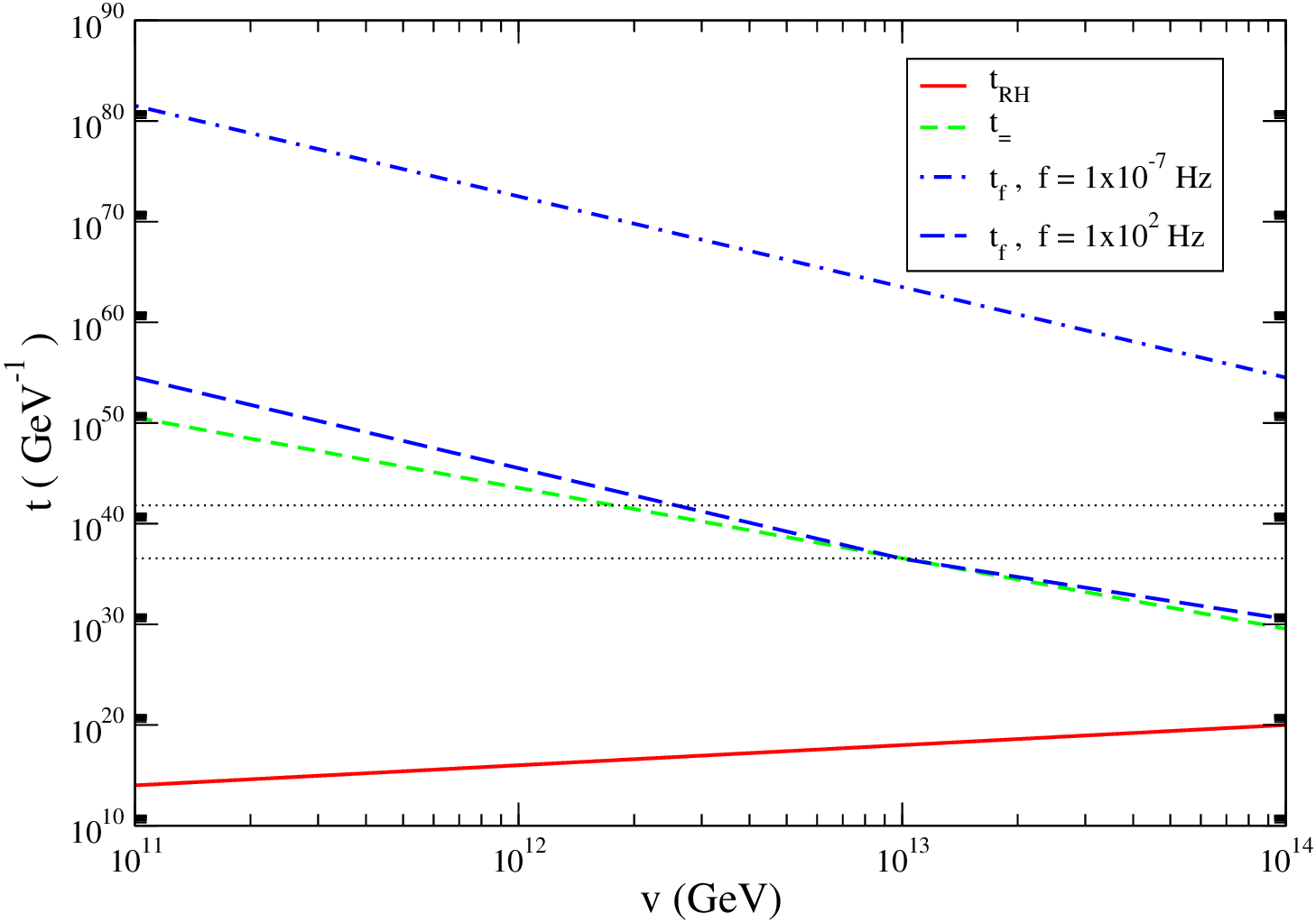}
\end{center}
\caption{Dependence of the times of interest $t_{RH}$, $t_=$,
and $t_f$ on the VEV $v$ for the representative small initial loop size
parameter $\alpha = 0.6\,\Gamma\,(G\mu)^{1.5}$.  
The black dotted lines indicate the present time 
$t_0 \simeq 6.6\times 10^{41}\,\gev^{-1}$
and the matter-radiation equality time 
$t_{eq}\simeq 3.5\times 10^{36}\,\gev^{-1}$.
}
\label{times-small}
\end{figure}

  The values of $t_{RH}$, $t_=$, and $t_f$ for very small loops,
$\alpha = 0.6\,\Gamma\,(G\mu)^{1.5}$, are shown in Fig.~\ref{times-small}
as a function of the VEV $v$.  This value of $\alpha$ corresponds to 
the estimate of Ref.~\cite{Polchinski:2006ee} 
for loops emitted in the matter era.
Even smaller values of $\alpha$ are suggested in Ref.~\cite{Siemens:2002dj}.
As before, the other model parameters were taken to be 
$m=10^3\,\gev$, $c=1$, $\gamma = 0.1$, and $\Gamma = 50$,
and the dotted lines denote the present time 
$t_0 \simeq 6.6\times 10^{41}\,\gev$ and the 
matter-radiation equality time $t_{eq} \simeq 3.5\times 10^{36}\,\gev$.
This figure indicates that the prospects for gravitational radiation
from small flat-direction string loops are much less promising than
for large loops.  Indeed, $t_=$ is larger than the present time $t_0$ 
for $v\lesssim 2\times 10^{12}\,\gev$.  
Such small loops will decay almost entirely to particles instead 
of gravitational radiation.  Even when $t_=$ is less than $t_0$,
the curves for $t_f$ show that the gravitational wave signal 
is very suppressed relative to the signal from large loops.
At low frequencies $f\simeq 10^{-7}\,Hz$ there is no signal at all 
since $t_f$ exceeds the present time $t_0$; the loops are simply
too small to radiate into this frequency range.  Even for frequencies
near $f=10^2\,Hz$, there will be a gravitational wave signal only for 
$v\gtrsim 3\times 10^{12}\,\gev$.  Despite the reduction in the gravitational
wave signal, small loops may be observable through their copious emission
of particles.

  In summary, we find that the large width of flat-direction cosmic
strings greatly enhances the rate at which they decay into their constituent
particles through cusp annihilation.  With this enhancement, 
our preliminary analysis indicates that string loops that are initially
large ($\alpha \simeq 0.1$) decay predominantly into gravitational waves, 
while very small loops ($\alpha \ll \Gamma\,G\mu$) 
decay primarily into particles.  The typical size of string loops
when they are formed is an unresolved problem, and well-motivated arguments
in favor of large loops, very small loops, or possibly both at once,
can be found in the literature.  In the face of this uncertainty,
we will focus on two particular choices of the loop size parameter $\alpha$
to estimate the observational signatures from flat-direction cosmic strings.
To compute the gravitational wave signals we will set $\alpha = 0.1$ 
for all loops, as suggested in Ref.~\cite{Vanchurin:2005yb}.  
Our results can be rescaled appropriately when only a fraction 
of the loops are large.  To estimate the signals from particle 
emission due to cusp annihilation, we will instead assume that $\alpha$ 
is sufficiently small that all loops decay mostly into particles.
This is plausible for flat-direction strings for which the rate
of particle emission by cusp annihilation is enhanced.
Again, it is straightforward to modify our results to accommodate 
larger values of $\alpha$.

  Finally, let us also mention that the picture of loop formation 
by flat-direction strings might be different from that of ordinary 
cosmic strings.  For example, the enhanced rate of particle emission 
by cusp annihilation could potentially smooth out small fluctuations over 
scales larger than the (na\"ive) gravitational radiation scale 
$\Gamma G\mu\,t$.  String loops can also remain bound to the parent 
string, as illustrated in Fig.~\ref{selfloop}.  This could modify 
the distribution of initial string loop sizes.  The rate
of cusp formation on these bound loops may also be different from
that on free loops.

\subsection{Gravitational Wave Signatures}

  Cosmic strings can give rise to two types of gravitational
wave signals.  The combination of many string loop decays 
produces a smooth \emph{stochastic} background of gravitational 
radiation~\cite{Vilenkin:1981bx}.  
On top of this background, individual cusps can produce 
intense bursts of gravity waves~\cite{Damour:2001bk}.
Gravitational wave detectors are sensitive to both types of signals.
For the string tensions of interest, $G\mu \lesssim 10^{-10}$,
the stochastic background is the more promising one~\cite{Siemens:2006vk,
Hogan:2006we,Siemens:2006yp} and we will focus on it.
To estimate this gravitational wave background due to
flat-direction cosmic strings we will assume that all
string loops are large when they are formed, 
with $\alpha \simeq 0.1$~\cite{Vanchurin:2005yb}.
If only a fraction of the loops produced are large, as advocated 
in Ref.~\cite{Polchinski:2007qc}, our results can be rescaled by
this fraction.

  We compute the gravitational radiation density due to cosmic string
decays following Ref.~\cite{Vachaspati:1984gt}.  
Consider radiation in the frequency range $(f,f+df)$ observed today 
that was emitted at time $\ttt$.  
Keeping track of only the lowest mode,\footnote{In Ref.~\cite{DePies:2007bm}
this was shown to be a good approximation for computing the 
stochastic background.} this radiation was emitted
by loops of size $(\tilde{\ell}-d\tilde{\ell},\tilde{\ell})$, where
\beq
\tilde{\ell} = \frac{2}{f}\frac{a(\ttt)}{a(t_0)},~~~~~
d\tilde{\ell} = \frac{2}{f^2}\frac{a(\ttt)}{a(t_0)}\,df.
\label{ltilde}
\eeq
Loops of this size at time $\ttt$ were formed at the earlier time 
$t_i$ given by
\beq
t_i = \lrf{1}{\alpha+\Gamma G\mu}\left[
\tilde{\ell}+\Gamma G\mu\tilde{t}\right],
\label{ti}
\eeq
over the time range
\beq
dt_i = \lrf{1}{\alpha+\Gamma G\mu}\frac{2}{f^2}\frac{a(\ttt)}{a(t_0)}\,df.
\eeq
These relations follow from the loop evolution equation 
${\ell}(t) =  \alpha\,t_i - \Gamma G\mu(t-t_i)$, valid for $t\geq t_i$
and $\ell (t)\geq \ell_=$.

  The rate at which loops are formed during the string scaling regime
can be estimated using the results of numerical simulations or from
simple analytic models like the one presented in Section~\ref{ntwrk}.
These predict a net energy flux into loops of
\beq
\frac{d\rho_{loop}}{dt} \simeq \frac{\rho_{\infty}}{t},
\label{loopdump}
\eeq
where 
\beq
\rho_{\infty} \simeq \zeta\mu\,t^{-2},
\eeq
with $\zeta\simeq 10$, and $\rho_{\infty}$ being the 
scaling energy density of long strings.  
This result can be obtained by summing
Eq.~\eqref{mastereq} over all string species that 
have equilibrated.  It follows that the rate per unit
volume that loops of initial size $\alpha\,t$ are formed is
\beq
\frac{d n}{dt}\simeq \frac{\zeta}{\alpha}\,t^{-4}.
\eeq

  Applying this result to loops formed in 
the time range $(t_i-dt_i,t_i)$, 
the number density of loops radiating into the 
frequency range of interest at time $\ttt$ is
\beq
dn(\ttt) \simeq \frac{\zeta}{\alpha}\,t_i^{-4}dt_i
\left[\frac{a(t_i)}{a(\ttt)}\right]^3.
\eeq
The redshift factor in this expression accounts for the dilution
of the loops as they evolve from $t_i$ to $\ttt$.  Given that
each loop radiates gravity waves with a power $\Gamma G\mu^2$,
we can combine everything and sum over $\ttt$ to find the signal.
The total gravitational wave density at the present 
frequency $f$ is
\bea
\Omega_{GW}(f) &:=& \frac{f}{\rho_c}\frac{d\rho_{GW}}{df}
\nnmb\\
&=& \frac{1}{\rho_c}\int_{\bar{t}_f}^{t_0}d\ttt\;
\Theta(\tilde{\ell}-\ell_=)\;
\Gamma G\mu^2\,f\frac{dn(\ttt)}{df}\,\left[\frac{a(\ttt)}{a(t_0)}\right]^4
\label{omegagw}\\
&\simeq& 
\frac{2}{f}\frac{\Gamma G\mu^2}{\rho_c}
\frac{\zeta}{\alpha(\alpha+\Gamma G\mu)}
\int_{\bar{t}_f}^{t_0}d\ttt\;\Theta(\tilde\ell-\ell_=)\;
\left[\frac{a(\ttt)}{a(t_0)}\right]^5\left[\frac{a(t_i)}{a(\ttt)}\right]^3\,
t_i^{-4}.\nnmb
\eea
Here, $\rho_c$ is the critical density, and $t_i$ and $\tilde{\ell}$
are functions of $\tilde{t}$ defined by Eqs.~\eqref{ltilde} and \eqref{ti}.
The integration limits range between $\bar{t}_f := max(t_f,10^{5}\,\gev^{-1})$ 
and $t_0$, where $t_f$ is given in Eq.~\eqref{tff}.\footnote{Normally the 
lower limit would simply be $t_f$, but in the present case the 
flat-direction string network only reaches scaling at 
$t\simeq 10^{5}\,\gev^{-1}$.  
Numerically, we find that this additional cutoff has no visible 
effect because the gravity waves emitted shortly after the
end of thermal inflation are diluted away during the subsequent reheating. 
Gravity waves from the phase transition~\cite{Grojean:2006bp} 
will also be diluted by thermal inflation.}
Noting that $a\propto t^{2/3}$ during the matter
era ($t<t_{RH}$ and $t>t_{eq}$) and $a\propto t^{1/2}$ during 
the radiation era ($t_{RH} < t < t_{eq}$), this equation can be
integrated straightforwardly.  
Relative to the treatment of Ref.~\cite{Vachaspati:1984gt}, 
we have included a cutoff of $\tilde{\ell} > \ell_==\alpha\,t_=$ 
through a step function.
This accounts for the loops only being able to radiate efficiently into gravity
waves if their length is greater than $\ell_=$.  It is this
cutoff, along with the additional redshifting that occurs during
reheating after thermal inflation, that suppresses the gravitational 
wave signal from flat-direction strings compared to ordinary strings.

\begin{figure}[ttt!]
\vspace{1.5cm}
\begin{center}
        \includegraphics[width = 0.75\textwidth]{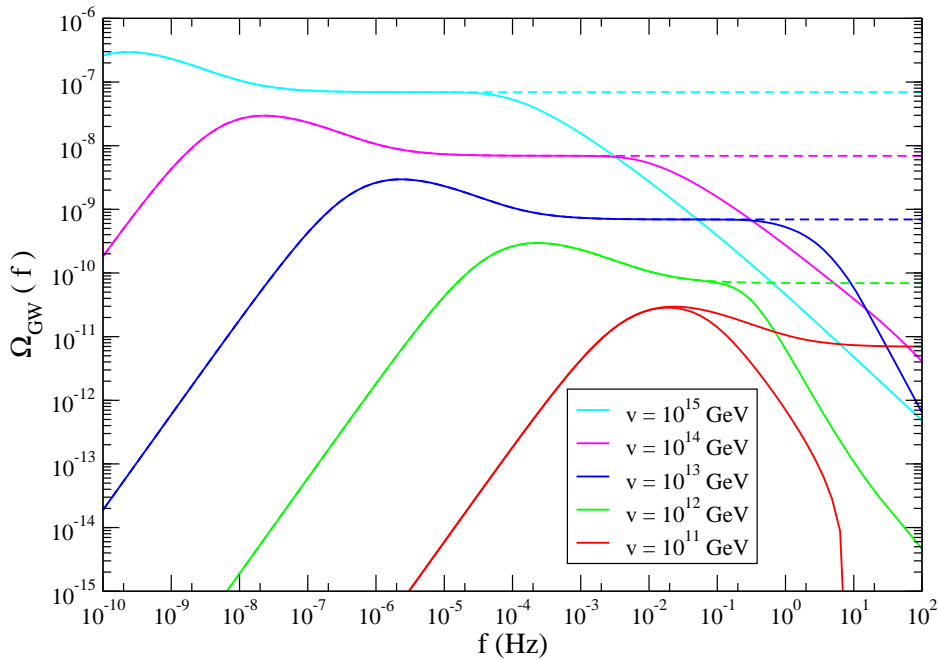}
\end{center}
\caption{Gravitational wave density for flat-direction cosmic strings
as a function of frequency for four different values of the VEV $v$.  
The solid lines include the cutoff
$\tilde{\ell}> \ell_=$ due to cusp annihilation.  The dashed lines
show what the gravitational wave density would be without this cutoff.}
\label{omegagwplot}
\end{figure}

  In Fig.~\ref{omegagwplot} we show the stochastic gravitational wave 
signal from initially large cosmic string loops as a function of frequency.  
We have used the parameter values $\alpha = 0.1$, $\Gamma = 50$,
$m=10^3\,\gev$, and $\gamma = 0.1$ in making this plot.
The solid lines show the gravitational wave density from flat-direction cosmic
strings, including the cutoff $\tilde{\ell} > \ell_{=}$ and the additional
redshifting during reheating after thermal inflation.  The dashed lines
indicate what the signal would be for ordinary cosmic strings,
without the cutoff $\tilde{\ell} > \ell_=$ or reheating effects.
At lower frequencies the relevant loops are formed later on,
at times greater than $t_{RH}$ and $t_=$, and there is no change
to the signal.  At higher frequencies, the cutoff on the loop size and
the additional dilution during reheating both suppress the gravitational
wave signals.  As can be seen in Fig.~\ref{times-big}, the cutoff
$\ell > \ell_==\alpha\,t_=$ is more important for lower values 
of $v$ (and $G\mu$), while the reheating dilution is more 
significant at larger values of $v$ since $t_{RH}$ is larger.  
This is why the shape of the high frequency cutoff changes 
as we increase $v$.  

  The attenuation of high frequency gravitational wave signals 
is relevant to LIGO and Advanced LIGO, 
which can potentially probe down to $\Omega_{GW}(f) \simeq 10^{-9}$ 
at frequencies around $f = 10^2\,Hz$~\cite{Abbott:2006zx}.  
Fig.~\ref{omegagwplot} indicates that LIGO is not expected to be able 
to find evidence for flat-direction cosmic strings.  On the other hand, 
the prospects for discovery at LISA and from measurements of 
pulsar timing are quite promising.  The LISA probe is expected 
to cover portions of the range 
$10^{-4}\,Hz \lesssim f \lesssim 10^{-2}\,Hz$ down to 
$\Omega_{GW}(f) \simeq 10^{-11}$~\cite{lisa}.  Since the gravitational wave
signal from large flat-direction string loops is mostly 
unmodified in this frequency range, LISA will be able to probe a 
sizeable portion of the model parameter space.
Limits from pulsar timing are currently 
$\Omega_{GW}(f) \lesssim 3\times 10^{-8}$ in the frequency
range $10^{-7}\!-\!10^{-8}\,Hz$~\cite{Jenet:2006sv}, 
which is again low enough that the gravitational wave signal 
from flat-direction cosmic strings is unsuppressed.
From this bound we obtain the constraint $v\lesssim 10^{14}\,\gev$.  
It is expected that this limit will be improved to 
$\Omega_{GW}(f) \lesssim 10^{-10}$ by upcoming experiments~\cite{Jenet:2006sv}.
Note that flat-direction cosmic strings offer the interesting 
possibility that LISA and pulsar timing experiments 
could detect a stochastic gravitational 
wave background with $\Omega_{GW}(f) \gtrsim 10^{-9}$, 
while (Advanced) LIGO sees nothing even though it is sensitive
to signals at this level.  This would be a suggestive hint
for flat-direction cosmic strings.

\subsection{Particle Emission Signatures: Dark Matter}

  Having studied the gravitational wave signatures of string
loops that are large when they are formed, let us now consider 
the possibility that the typical initial loop size is very small, 
$\alpha \ll \Gamma\,G\mu$, as suggested in Refs.~\cite{Siemens:2002dj,
Polchinski:2006ee}.
If $\alpha$ is sufficiently small, nearly all the energy of a loop 
is released as particle excitations of the fields making up the string.
This is plausible for flat-direction cosmic strings due to
their enhanced rate of particle emission by cusp annihilation 
relative to ordinary cosmic strings.
In the $(a,b)$ models of flat-direction strings presented in
Section~\ref{internal}, the fields making up the string 
consist of two chiral supermultiplets and one massless vector (gauge) 
supermultiplet.  When the chiral supermultiplets develop VEVs, 
it is more convenient to describe the theory in terms of a heavy massive 
vector supermultiplet, with mass on the order of $gv$, 
as well as a light supermultiplet with mass on the order 
of $m$~\cite{Morrissey:2006xn}.  
This light multiplet is light on account 
of the flatness of the potential.  Cusp annihilation will produce
both the heavy and the light states making up the string.
These particles will subsequently decay, and can be a potential source 
of dark matter and high energy cosmic rays.  
We consider both of these possible signatures in turn,
assuming that all string loops are very small and decay entirely
into particles rather than gravitational waves. 

  To compute the dark matter density from decaying string loops
we again make use of Eq.~\eqref{loopdump} that specifies 
the rate at which the scaling string network transfers its energy
into loops.  Contributions to the dark matter density from loops
produced before the network attains scaling are diluted away
by the subsequent reheating process.\footnote{We have
verified this using the simulation of Section~\ref{ntwrk}.}
We assume that a fraction $\epsilon_1$ of the energy
emitted by the cusp annihilations of loops eventually becomes 
dark matter (such as a neutralino or gravitino LSP).\footnote{
The decay products from a cusp annihilation are typically
boosted by factor of $(\ell/w)^{1/2}$~\cite{BlancoPillado:1998bv}.
This can lead to values of $\epsilon_1 \ll 1$ if most of the
cusp energy goes to kinetic energy rather than non-relativistic
dark matter.  For small initial loop sizes, $\alpha < \Gamma G\mu$, 
this boost is weak enough that the dark matter decay products 
quickly reach kinetic equilibrium for $T \gtrsim T_{RH}$, 
when most of the dark matter is produced.}
The total dark matter density at the present time from the strings is then
\beq
\rho_{DM}^{strings} \simeq 
\epsilon_1\int^{t_0}_{t_{fo}}\,dt\;
\zeta\,\frac{\mu}{t^3}\left[\frac{a(t)}{a_0}\right]^3,
\eeq
where $t_0$ is the present time and $t_{fo}$ is the time 
at which the DM particles freeze out of equilibrium.  
The factor of $[a(t)/a_0]^3$ accounts for the additional dilution
of the dark matter (or the constituent string fields) 
after they are produced.
It is convenient to split the integration into three pieces: 
$t_{eq} < t$, $t_{RH} < t < t_{eq}$, and $t_{fo} < t < t_{RH}$.  
These integrations are straightforward and yield
\bea
\Omega_{DM}^{strings} &\simeq& 
6\pi\epsilon_1\zeta\,G\mu\left[
\ln\lrf{t_0}{t_{eq}} 
+ \lrf{t_{eq}}{t_{RH}}^{1/2}
+ \ln\lrf{t_{RH}}{t_{fo}}\,\lrf{t_{eq}}{t_{RH}}^{1/2}\right]\\
&\lesssim& 30\;\epsilon_1\,\lrf{\gamma}{0.1}^{1/2}\lrf{\zeta}{10}
\lrf{v}{10^{14}\,\gev}\lrf{m}{10^3\,\gev}^{3/2},\nnmb
\eea
where $\gamma$ is the prefactor appearing in Eq.~\eqref{decayrate}.
Numerically, the largest contribution comes from
the third term, from the integration range $t_{fo} < t < t_{RH}$.
We have bounded the logarithm in this term from above in making this
estimate.  For reasonable values of the model parameters,
the amount of dark matter produced by decaying loops is safely small,
although smaller values of $v$ are preferred.
This differs from the much stronger constraints on regular cosmic strings 
that are able to decay into dark matter~\cite{Jeannerot:1999yn}, 
which is due to the dilution from reheating after thermal inflation.

  There is an additional contribution to the DM density from
the out-of-equilibrium decays of the oscillating flat-direction fields
during reheating.  If a small fraction $\epsilon_2$ of these decays 
ends up as dark matter, the present contribution to the DM energy 
density will be
\beq
\rho_{DM}^{\phi} \simeq \int_{t_{fo}}^{t_{RH}}dt\,\epsilon_2\,
\Gamma_{\phi}\,\rho_{\phi}\,\left[\frac{a(t)}{a_0}\right]^3.
\eeq
Parametrizing $\rho_{\phi}(t) \simeq T_{RH}^4[a(t_{RH})/a(t)]^3$
and using $t_{RH}\Gamma_{\phi} \simeq 1$, we find
\beq
\Omega_{DM}^{\phi} \simeq \,10^7\;\epsilon_2\,
\lrf{T_{RH}}{\gev}.
\eeq
Thus, the branching fraction $\epsilon_2$ 
into DM particles must be very small.
Note that $\epsilon_1$ and $\epsilon_2$ can be very different from
each other.  The particles emitted from a cusp annihilation can include
some of the heavier component fields making up the string.  On the other hand,
the decays of the flat-direction fields after thermal inflation 
involve only the light modes.  The decays of these states into superpartners
(such as a neutralino or heavier gravitino LSP) can therefore 
be highly suppressed or kinematically inaccessible, 
allowing for $\epsilon_2\ll\epsilon_1$.

\subsection{Particle Emission Signatures: Visible Matter}

  In addition to dark matter, the decays of very small string loops 
can produce hadrons, leptons, and photons.  This particle injection 
will be spread out over time as the scaling string network continually
rids itself of excess energy by forming loops.  Visible particles created 
by loop decays can imprint themselves upon the early universe 
in a number of ways.  The energetic products from loop decays at temperatures
below $5\,\mev$ can disrupt the predictions of big-bang
nucleosynthesis~(BBN).  At later times, energetic photons from 
loop decays can modify the blackbody spectrum 
of the cosmic microwave background~(CMB).  Some of the decay products
from string loops can also be highly energetic, producing 
ultra-high-energy cosmic rays and contributing 
to the extragalactic diffuse gamma-ray background.  We consider 
the possible signatures from cosmic strings from each of these effects.  
As for our dark matter estimates, we assume that all loops 
are so small that they decay entirely into particles.  

  To estimate the effects of decaying string loops on BBN
we make use of the results of Ref.~\cite{Kawasaki:2004qu}.  
In this work the authors used the successful predictions of BBN to place
limits on the combination $m_XY_X$ for a long-lived particle $X$ of mass 
$m_X$, relic density (per unit entropy) $Y_X$, and lifetime $\tau_X$, 
decaying at time $t\simeq \tau_X$.  
In contrast to a long-lived relic particle whose decays can be treated
as being instantaneous, cosmic string loops are produced
and decay continuously.  These decays therefore have a cumulative
effect on the light element abundances.  To obtain a limit for decaying
string loops we interpret the bounds from Ref.~\cite{Kawasaki:2004qu} as
limits on the total energy injected within a comoving
volume, $m_XY_X = \Delta E/S$, where $S$
is the total entropy within the volume $a^3$.  

  The total energy injected into the comoving volume $a^3$ by 
string loops that decay during the time interval $(t_a,t_b)$ is 
\bea
\frac{\Delta E}{S} &=& \frac{1}{S}\;\int_{t_a}^{t_b}dt\;
\mu\,\zeta\,t^{-3}\,{a^3(t)}\\
&\simeq& 
(10^{-11}\,\gev)\,
\lrf{1s}{t_a}^{1/2}\,\lrf{\zeta}{10}\,\lrf{G\mu}{2\times 10^{-11}}.
\nnmb
\eea
In writing this expression, we have implicitly assumed that $t_a$ and
$t_b$ both lie within the era of radiation dominance, as is relevant for BBN.  
The strongest limits on energy injection 
from BBN come from the relative fractions of deuterium and lithium-6 
relative to hydrogen. 
Both of these are formed at times later than $t\gtrsim 100\,s$.  
Since the visible decay products from the loops thermalize quickly
relative to the Hubble time, we set $t_a = 100\,s$ to find the
bounds due to the deuterium and lithium-6 abundances~\cite{Kawasaki:2004qu}.
Assuming a hadronic branching fraction 
of order unity, the total energy injection per 
unit entropy must be less than $\Delta E/S \lesssim 10^{-14}\,\gev$.\footnote{
The bound is fairly independent of the mass of the decaying particle.}
This bound is satisfied provided $v\lesssim 10^{13}\,\gev$.  

  Late-time energy injection is also constrained by the 
the nearly perfect blackbody spectrum of the CMB observed 
by COBE/FIRAS~~\cite{Fixsen:1996nj}.
Photons produced by the decays of string loops that occur after
the time $t_{dC} \simeq 10^{31}\,\gev^{-1}$ can distort this spectrum.
Before $t_{dC}$, double Compton scattering ($e+\gamma\to e+\gamma+\gamma$)
efficiently thermalizes any additional photons that are created.
The precise form of the spectral distortions created after $t_{dC}$ 
depends on the time at which the photons were injected.  
However, the net constraint from the non-observation of such distortions 
can be reduced to a constraint on the total photon energy created 
after $t_{dC}$, 
$\Delta\rho_{\gamma}/\rho_{\gamma} \lesssim 7\times 10^{-5}$~\cite{
Brandenberger:2002qh,Hu:1992dc,Hagiwara:2002fs}.
The net photon injection from decaying string loops can be estimated
using the rate of energy deposition by the network.  If all the energy
injected is in the form of photons (possibly after cascading), the 
total injection is
\bea
\frac{\Delta\rho_{\gamma}}{\rho_{\gamma}}(t_0) &\simeq&
\frac{1}{\rho_{\gamma_0}}\,
\int_{t_{dC}}^{t_0}dt\;\frac{\del\rho}{\del t}\left[
\frac{a(t)}{a(t_0)}\right]^4\\
&\simeq&
{6\pi}\,\zeta\,G\mu\;\left[\ln\lrf{t_{eq}}{t_{dC}} 
+ \frac{1}{\Omega_{\gamma_0}}\right]
\nnmb\\
&\simeq& (8\times 10^{-5})\lrf{\zeta}{10}\lrf{G\mu}{2\times 10^{-11}}.
\nnmb
\eea
Numerically, the dominant contribution to the injected photon energy comes
from the most recent era, $t>t_{eq}$, leading to a non-zero value
for the Compton $y$ parameter~\cite{Hu:1992dc,Kanzaki:2007pd} which quantifies
deviations away from the black body spectrum.  
It is expected that the constraints
on photon injection will be improved in the future by the ARCADE 
experiment~\cite{Kogut:2006kf}.

  Decaying cosmic string loops can also generate cosmic rays.
The corresponding energy spectrum depends on the energies of
the particles emitted in the loop decays.  Recall that the
fields making up the flat-direction strings consist 
of a light chiral supermultiplet and a heavy massive 
vector supermultiplet.  In each cusp annihilation,
both the heavy and the light states can be produced.
The decays of the heavy states, with masses on the order of $g\,v$, 
can generate ultra-high-energy cosmic rays~(UHECR)~\cite{cosmray}.  
Decays of the light states, with masses on the order of $m \ll v$, 
contribute to the extragalactic diffuse gamma-ray
background~(EDGRB)~\cite{Bhattacharjee:1997in}.  
To determine the relevant bounds and prospects, 
we will assume that the energy released in each cusp annihilation 
goes initially into a fraction $F_l$ of the light states (with soft energy)
and a fraction $F_h$ to the heavy states.  We expect $F_l\sim 1$, 
with $F_h$ possibly smaller.

    The contribution of decaying string loops 
to the EDGRB was studied in Ref.~\cite{Bhattacharjee:1997in}.  
Data from the EGRET experiment~\cite{Sreekumar:1997un} constrains the rate 
of energy emission into the light scalar states with masses on the order
of $1000\,\gev$ (that decay into lower-energy gamma rays) at the present
time to $\partial \rho_{loop}/\partial t_0 
\lesssim 4.5\times 10^{-23}\,\mbox{eV}\,cm^{-3}\,s^{-1}
= 2.3\times 10^{-97}\,\gev^5$.
Equating this bound with Eq.~\eqref{loopdump} evaluated at the present time,
we obtain the bound~\cite{Bhattacharjee:1997in}
\beq
F_l\lrf{\zeta}{10}\lrf{G\mu}{2\times 10^{-11}} \lesssim 1.
\label{edgrb}
\eeq
This does not represent a significant constraint beyond those
found above.  The heavy component states making up
the string can also contribute to the EDGRB through the photons
they produce in cascade decays.  The limit in this case is 
about the same as from the decays of the light states given in
Eq.\eqref{edgrb}, but with $F_l$ replaced by $F_h$.
These constraints from the gamma-ray background on decaying cosmic string
loops will be strengthened by the upcoming GLAST 
experiment~\cite{Bloom:1994eu}.  
However, the range of the model that can be probed may 
ultimately be limited by astrophysical background 
contributions to the gamma-ray flux.

  Ultra-high-energy cosmic rays can be produced 
by cusp annihilation if some of the heavier states making up 
the string are created.  When the heavy states decay, 
their products are highly energetic, making them a source of 
high-energy neutrinos and UHECRs.  
Estimates of the UHECR flux for strings that decay into particles 
were made in Ref.~\cite{Wichoski:1998kh}
and are directly applicable to flat-direction cosmic strings.  These
authors find that for energies greater than about $6\times 10^{9}\,\gev$, 
the only relevant cosmic ray flux consists of neutrinos.  
The fluxes of highly energetic protons and photons 
are very suppressed because they are attenuated by their 
interactions with the cosmic background.
Extrapolating the predictions of Ref.~\cite{Wichoski:1998kh}, 
the neutrino signal from decaying strings can be probed directly at 
Ice Cube~\cite{Filimonov:2006zs} down to $G\mu\lesssim 10^{-12}/F_h$
in the energy range $10^5\,\gev$-$10^{8}\,\gev$.
The Auger project is sensitive to UHECR showers induced by energetic
neutrinos in the energy range 
$10^{9}\,\gev$-$10^{11}\gev$~\cite{Facal San Luis:2007it}. 
The Auger measurements imply the constraint
\beq
G\mu \lesssim (3\times 10^{-13})/F_h.
\eeq
For $F_h = 1$, this corresponds to $v\lesssim 10^{13}\,\gev$.

  Our analysis indicates that the visible matter signatures
from decays of flat-direction string loops are consistent
with observations provided $G\mu$ is small enough.
However, there is another visible matter signature that
is challenging to reproduce in models of flat-direction strings,
namely the baryon asymmetry of the universe.  Flat-direction
strings are formed following a period of thermal inflation.
The typically low reheating temperature after thermal inflation,
Eq.~\eqref{trh}, combined with the large amount of dilution from the 
inflationary expansion and reheating imply that baryogenesis mechanisms
that operate at or above the electroweak scale will no longer work.
Instead, the baryon asymmetry must be produced at very late times.  
This can arise from the strings themselves~\cite{Kawasaki:1987vg,
Mohazzab:1994xj,Dasgupta:1996ys}, 
from the non-thermal production of particles during reheating 
that have baryon-number violating 
decays~\cite{Yamamoto:1986jw,Lazarides:1986di,
Dimopoulos:1987rk,Cline:1990bw,Babu:2006xc}, 
or by the Affleck-Dine mechanism~\cite{Stewart:1996ai}.

\subsection{String Loops and Zero Modes}

  In our discussion of radiation from cosmic string loops,
we implicitly assumed that there do not exist any
\emph{zero mode} excitations along the strings.
Zero modes are fermionic or bosonic field fluctuations with vanishing
energy that are localized on the string.  The existence of zero modes
on cosmic strings can alter the picture of loop radiation
in important ways~\cite{Witten:1984eb,Carter:1990sm,
Brandenberger:1996zp,Davis:2003an}.

  These undamped, particle-like excitations can be excited when a string loop
is formed.  As the loop radiates and shrinks, the number density of
the zero modes builds up.  Eventually the angular momentum of the zero
modes balances the tendency of the loop to shrink, and a quasi-stable
loop remnant, or \emph{vorton}, is left over.  If such vortons are
sufficiently long-lived and numerous, they behave
like quasi-stable matter and can further modify the predictions of BBN
or create too much dark matter.  The presence of vortons typically leads
to extremely strong constraints on the underlying
field theory~\cite{Davis:2003an}.

  For the flat-direction strings we are studying, fermionic zero
modes~\cite{Jackiw:1981ee,Witten:1984eb} are of particular relevance.
It was shown in Ref.~\cite{Davis:1997bs,Davis:1997ny} 
that such modes are a generic
feature of supersymmetric cosmic string solutions.
In the present case, we also have supersymmetry
breaking operators present in the Lagrangian.  We find that adding
a supersymmetry breaking gaugino mass destroys all the fermionic
zero modes.  A recent study also suggests that more generally,
fermionic zero modes do not form on closed string loops
at all~\cite{Postma:2007bf}.  The existence of bosonic zero modes 
depends on the other fields in the theory and their couplings, 
and are less generic~\cite{Witten:1984eb}.  We do not consider them here.  

  Zero modes, either bosonic or fermionic, are also unlikely to
stabilize flat-direction cosmic strings simply because these strings
are relatively wide.  
For the phase transition leading to flat-direction cosmic strings, 
we expect that the radius at which the zero modes would stabilize 
a string loop, if they were to exist, 
is usually much smaller than the width of the string~\cite{Davis:2005jf}.  
As discussed above for cusp annihilation, when the 
separation between a pair of antiparallel string segments approaches
the string width these segments will annihilate into their constituent
fields, and the loop will decay before stabilizing as a vorton.

\subsection{Lensing by Cosmic Strings}

  While an indirect gamma ray or gravitational wave signal from
cosmic strings would be exciting, ideally one would like a direct
observation to confirm their existence.  This can be achieved by
observing gravitational lensing by a string.  The primary gravitational 
effect of the large mass density contained within a cosmic string is 
to modify the surrounding spacetime such that it is flat, but with a deficit
angle of $\Delta \theta = 8\pi\,G\mu$~\cite{Vilenkin:lens}.  
When light from a galaxy passes by a (non-relativistic) cosmic string, 
the deficit angle produces a distinctive double image with an angular 
separation of~\cite{sbook,Vilenkin:lens}
\beq
\Delta \alpha = 8\pi\,G\mu\,\frac{D_{ls}}{D_{os}}\,\sin\phi,
\eeq
where $D_{ls}$ is the distance from the lensing string to the source galaxy,
$D_{os}$ is the distance from the observer to the source, and $\sin\phi$
is the angle between the string axis and the line-of-sight.
From a single lensing event it is possible to determine $\Delta \alpha$
directly, as well as $D_{os}$ by measuring the redshift of the source.  
Given that a single string lensing event is found, it is likely that
the same string will also lens the images of other galaxies that are nearby
on the sky~\cite{Huterer:2003ze}.  By observing several galaxies lensed 
by the same string, the tension of that string can be 
determined~\cite{Oguri:2005dt}.

  The gravitational lensing signatures from flat-direction strings are even
richer than those of ordinary strings because of the stability of 
higher winding modes.  If many lensed images from different strings
are observed, it may be possible to measure tensions of several strings
and obtain clues about the mass spectrum of the higher winding modes.  
In this respect, flat-direction cosmic strings are similar to $(p,q)$ 
cosmic superstrings.  
Both types of cosmic strings also have junctions connecting different 
winding modes.  These can produce triple images, in addition to the 
double images produced by a lone string~\cite{Shlaer:2005ry}.  
Since the spectrum of tensions of flat-direction strings is very 
different from that of $(p,q)$ strings, the observation of many 
gravitational lensing events might allow one to distinguish 
between them.  
Unfortunately, the probability of observing a lensed image from
a flat-direction string is very small due the indirect bounds on
the tension, $G\mu \lesssim 10^{-11}$.  This is smaller than the expected
sensitivity of $G\mu \simeq 10^{-8}$ from upcoming optical 
surveys~\cite{Gasperini},
and $G\mu \simeq 10^{-9}$ from the SKA~\cite{Diamond:2006wb}, 
radio survey~\cite{Mack:2007ae}.
The results of Ref.~\cite{Oguri:2005dt} also suggest that it will
be difficult to determine the string tension accurately enough
to distinguish between flat-direction string winding modes with similar
values of the winding number $N$.

\section{Conclusion\label{disc}}

We conclude by summarizing some of our findings:

\begin{itemize}

\item Abelian gauge symmetry breaking along a flat direction 
can give rise to strongly Type-I
cosmic strings with tension $\mu\simeq 0.1\pi v^2$, gauge profile width 
of $v^{-1}$ and scalar profile width $w \sim m^{-1}$, where $m\ll v$ 
characterizes the flatness of the potential. 
These flat-direction strings are likely to be formed after 
thermal inflation through flux-trapping.

\item The tension of the strings increases very slowly with 
their winding number $N$. Thus, higher-winding mode strings 
$N=2,3,\ldots$ are energetically stable. This enables
strings to be attracted to one another and zipper, creating 
stable formations with winding number $N_1+N_2$ or $|N_1-N_2|$, 
where $N_1$ and $N_2$ are the original string winding numbers.

\item Zippering affects the evolution of 
the resulting string network.  Applying a simple network 
evolution model to flat-direction strings 
suggests that a large number of string modes
develop roughly equal densities in the early universe.
The total energy density is about the same as for a single string,
but it is distributed among many species.

\item Flat-direction strings radiate gravitationally.
However, in contrast to ordinary cosmic strings, 
they also may be able to radiate copiously into matter.  
The strings are expected to fully radiate away, 
as there is no vorton obstruction for the 
supersymmetric flat-direction strings under consideration. 

\item In contrast to GUT strings, flat-direction strings are 
generically compatible with current direct observational constraints,
$G\mu \lesssim 3\times 10^{-7}$~\cite{
Wyman:2005tu,Fraisse:2006xc,Pogosian:2006hg,Seljak:2006bg,Bevis:2006mj}.  
If the typical initial loop length is close to the horizon scale, 
LISA and upcoming millipulsar timing
probes may be able to detect the gravitational wave signal from
these strings.  However, the gravity wave signal at higher frequencies 
is suppressed for flat-direction strings, making their detection at
LIGO extremely challenging.

\item Particle emission from cusp annihilation is likely to be the
dominant loop decay mechanism if the loop length is always much
smaller than the horizon.  This intriguing prospect for flat-direction
cosmic strings entertains the possibility 
that ultra-high-energy cosmic rays or nonthermal dark matter 
originate from their particle emission.  
If all loops decay entirely into particles, the constraints from
BBN, the CMB blackbody, and UHECRs imply the bound $v\lesssim 10^{13}\,\gev$,
corresponding to $G\mu \lesssim 10^{-13}$.

\item The multi-tension network of flat-direction strings 
formed in the early universe is in contrast to the standard 
single-tension string networks, but similar to $(p,q)$ 
cosmic superstring networks, 
and thus may mimic the latter by giving rise to multiple 
lensing events.  However, the spectrum of tensions of 
flat-direction strings is constrained by indirect bounds,
and may be too low to be observed in the near future.

\end{itemize}

  We find that flat-direction cosmic strings behave in ways that are
qualitatively different from both ordinary (abelian Higgs) cosmic
strings as well as $(p,q)$ cosmic superstrings.  These differences
in behavior may be distinguishable through probes of the early universe.

\section*{Acknowledgements}
We thank L. Bettencourt, D. Chung, A. Pierce, E. Rozo, and K. Turzynski 
for helpful conversations, and N. Arkani-Hamed for pushing us to look
more closely at the boosted decay products from cusps. 
This work was supported by the National Science Foundation under 
Grant No.~PHY-0456635, the Department of Energy, 
and the Michigan Center for Theoretical Physics (MCTP).


\end{document}